\documentstyle [12pt,eqsecnum,aps,amsfonts] {revtex}
\input epsf
\newcommand{\beq}{\begin{equation}}
\newcommand{\eeq}{\end{equation}}
\newcommand{\beqs}{\begin{eqnarray}}
\newcommand{\eeqs}{\end{eqnarray}}

\begin{document}
\draft

\baselineskip 6.0mm

\title{Ground State Entropy of Potts Antiferromagnets on Cyclic 
Polygon Chain Graphs} 

\author{Robert Shrock$^{(a)}$\thanks{email: robert.shrock@sunysb.edu}
\and Shan-Ho Tsai$^{(b)}$\thanks{email: tsai@hal.physast.uga.edu}}

\address{(a) \ Institute for Theoretical Physics \\
State University of New York \\
Stony Brook, N. Y. 11794-3840}

\address{(b) \ Department of Physics and Astronomy \\
University of Georgia \\
Athens, GA  30602}

\maketitle

\vspace{10mm}

\begin{abstract}

We present exact calculations of chromatic polynomials for families of cyclic
graphs consisting of linked polygons, where the polygons may be adjacent or
separated by a given number of bonds.  From these we calculate the 
(exponential of the) ground state entropy, $W$, for the $q$-state Potts model
on these graphs in the limit of infinitely many vertices.  A number of
properties are proved concerning the continuous locus, ${\cal B}$, of 
nonanalyticities in $W$.  Our results provide further evidence for a 
general rule concerning the maximal region in the complex $q$ plane to which
one can analytically continue from the physical interval where $S_0 > 0$.

\end{abstract}


\vspace{16mm}

\pagestyle{empty}
\newpage

\pagestyle{plain}
\pagenumbering{arabic}
\renewcommand{\thefootnote}{\arabic{footnote}}
\setcounter{footnote}{0}

\section{Introduction}

The $q$-state Potts antiferromagnet (AF) \cite{potts,wurev} exhibits nonzero
ground state entropy, $S_0 > 0$ (without frustration) for sufficiently large
$q$ on a given graph or lattice.  This is equivalent to a ground state
degeneracy per site $W > 1$, since $S_0 = k_B \ln W$.  Such nonzero ground
state entropy is important as an exception to the third law of thermodynamics
\cite{al}.  The zero-temperature partition function of the above-mentioned
$q$-state Potts antiferromagnet on a graph $G$ satisfies
$Z(G,q,T=0)_{PAF}=P(G,q)$, where $P(G,q)$ is the chromatic polynomial
expressing the number of ways of coloring the vertices of the graph $G$ with
$q$ colors such that no two adjacent vertices have the same color
\cite{bl}-\cite{rtrev}. Thus
\beq
W(\{G\},q) = \lim_{n \to \infty} P(G,q)^{1/n}
\label{w}
\eeq 
where $n=v(G)$ is the number of vertices of $G$ and $\{G\} = \lim_{n \to
\infty}G$. The minimum number of colors needed for this 
coloring of $G$ is called its chromatic number, $\chi(G)$. 
At certain special points $q_s$ (typically $q_s=0,1,.., \chi(G)$),
one has the noncommutativity of limits
\beq
\lim_{q \to q_s} \lim_{n \to \infty} P(G,q)^{1/n} \ne \lim_{n \to \infty} 
\lim_{q \to q_s}P(G,q)^{1/n}
\label{wnoncom}
\eeq 
and hence it is necessary to specify the order of the limits in the
definition of $W(\{G\},q_s)$ \cite{w}.  As in \cite{w}, we shall use the first
order of limits here; this has the advantage of removing certain isolated
discontinuities in $W$.  In addition to Refs. \cite{wurev}-\cite{w}, some
other previous related works include Refs. \cite{lieb}-\cite{pg}. 
Since $P(G,q)$ is a polynomial, one can generalize $q$
from ${\mathbb Z}_+$ to ${\mathbb R}$ and indeed to ${\mathbb C}$.
$W(\{G\},q)$ is a real analytic function for real $q$ down to a minimum value,
$q_c(\{G\})$ \cite{w,p3afhc}.  For a given $\{G\}$, we denote the continuous
locus of non-analyticities of $W$ as ${\cal B}$.  This locus ${\cal B}$ forms
as the accumulation set of the zeros of $P(G,q)$ (chromatic zeros of $G$) as $n
\to \infty$ \cite{bds,bkw,read91,w} and satisfies ${\cal B}(q)={\cal B}(q^*)$.
A fundamental question concerning the Potts antiferromagnet is the form of this
locus for a given graph family or lattice and, in particular, the maximal
region in the complex $q$ plane to which one can analytically continue the
function $W(\{G\},q)$ from physical values where there is nonzero ground state
entropy, i.e., $W > 1$.  We denote this region as $R_1$. Further, we denote as
$q_c$ the maximal point where ${\cal B}$ intersects the real axis, which can
occur via ${\cal B}$ crossing this axis or via a line segment of ${\cal B}$
lying along this axis.

In the present work we present exact calculations of
$P(G,q)$ and $W(\{G\},q)$ for families of cyclic graphs consisting of linked
polygons, where the polygons may be adjacent or separated by a certain number
of bonds.  By the term cyclic we mean that these families of graphs contain a
global circuit, defined as a route along the graph which has the topology of
$S^1$ and a length $\ell_{g.c.}$ that goes to infinity as $n \to \infty$.  The
results provide new insight into $W$ and ${\cal B}$ in an exactly solvable
context where one can study their dependence on several parameters
characterizing the families of polygon chain graphs.  The present work extends
our recent studies of cyclic families of graphs \cite{w,pg}.

  From our previous exact
calculations of $P(G,q)$ and $W(\{G\},q)$ on a number of families of graphs we
have inferred several general results on ${\cal B}$: (i) for a graph $G$ with
well-defined lattice directions, a sufficient condition for ${\cal B}$ to
separate the $q$ plane into different regions is that $G$ contains at least one
global circuit \cite{strip}\footnote{
Some families of graphs that do not have regular lattice
directions have noncompact loci ${\cal B}$ that separate the $q$ plane into
different regions \cite{wa,wa3,wa2}.}, and (ii) the locus ${\cal B}$ for such a
graph does not contain any endpoint singularities.  Two other general features
are that for graphs that (a) contain global circuits, (b) cannot be written in
the form $G=K_p + H$ \cite{wc} 
\footnote{The complete graph on $p$ vertices, denoted $K_p$, is the graph in
which every vertex is adjacent to every other vertex. The ``join'' of graphs
$G_1$ and $G_2$, denoted $G_1 + G_2$, is defined by adding bonds linking each
vertex of $G_1$ to each vertex in $G_2$.} and (c) have compact ${\cal B}$, we
have observed that ${\cal B}$ (iii) passes through $q=0$ and (iv) crosses the
positive real axis, thereby always defining a $q_c$.  The families of polygon
chain graphs studied in this paper do contain global circuits, and the exact
results presented here are in accord with the above general properties.

The chromatic polynomial $P(G,q)$ has a general decomposition as
\beq
P(G,q) = c_0(q) + \sum_{j=1}^{N_a} c_j(q)(a_j(q))^{t_j n}
\label{pgen}
\eeq 
where the $a_j(q)$ and $c_{j \ne 0}(q)$ are independent of $n$, while
$c_0(q)$ may contain $n$-dependent terms, such as $(-1)^n$, but does not grow
with $n$ like $(const.)^n$ with $|const.| > 1$.  The expression $c_0$ may be
absent.  A term $a_\ell(q)$ is ``leading'' if it dominates the $n \to \infty$
limit of $P(G,q)$.  The locus ${\cal B}$ occurs where there is an abrupt
nonanalytic change in $W$ as the leading terms $a_\ell$ changes; thus this
locus ${\cal B}$ is the solution to the equation of degeneracy of magnitudes of
leading terms.  Hence, $W$ is finite and continuous, although nonanalytic,
across ${\cal B}$.

It is convenient to define the following polynomial: 
\beq
D_k(q) = \frac{P(C_k,q)}{q(q-1)} = a^{k-2}\sum_{j=0}^{k-2}(-a)^{-j} =
\sum_{s=0}^{k-2}(-1)^s {{k-1}\choose {s}} q^{k-2-s}
\label{dk}
\eeq
where
\beq
a = q-1
\label{a}
\eeq
and $P(C_k,q)$ is the chromatic polynomial for the circuit
(cyclic) graph $C_k$ with $k$ vertices,
\beq
P(C_k,q) = a^k + (-1)^ka
\label{pck}
\eeq

This paper is organized as follows.  In Sections II and III we present our
results for $P$, $W$, and ${\cal B}$ for the open and cyclic polygon chain
graphs.  Some concluding remarks are included in Section IV, and some further
results are included in the Appendix. 

\section{Open Polygon Chain Graphs}

\begin{figure}
\centering
\leavevmode
\epsfxsize=4.0in
\epsffile{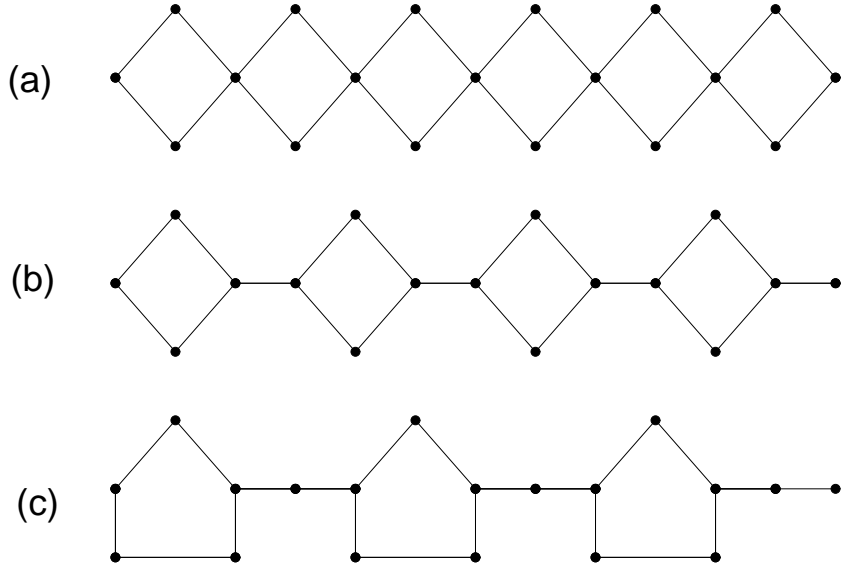}
\vspace{1 cm}
\caption{\footnotesize{Illustrations of cyclic and open polygon chain graphs
$G_{e_1,e_2,e_g,m}$ and $G_{e_1,e_2,e_g,m;o}$ with $(e_1,e_2,e_g,m)=$ (a) 
(2,2,0,6), (b) (2,2,1,4), (c) (2,3,2,3).  For the cyclic (open) graphs, the 
rightmost vertex on each graph is identified with (is distinct from) the 
leftmost vertex at the same level, respectively.}}
\label{pgchain}
\end{figure}

Before giving our results for the cyclic polygon chain graphs, we discuss the
simpler case of families of open polygon chain graphs comprised of $m$
subunits, each subunit consisting of a $p$-sided polygon with one of its
vertices attached to a line segment of length $e_g$ edges (bonds).  Thus, the
members of each successive pair of polygons are separated from each other by a
distance (gap) of $e_g$ bonds along these line segments, with $e_g=0$
representing the case of contiguous polygons.  Since each polygon is connected
to the rest of the chain at two vertices (taken to be at the same relative
positions on the polygons in all cases), the family of graphs depends on two
additional parameters, namely the number of edges of the polygons between these
two connection vertices, moving in opposite directions along the polygon, $e_1$
and $e_2$. For $e_1 \ne e_2$ we denote the smaller ($s$) and larger ($\ell$)
edge lengths as 
\beq
e_s \equiv min(e_1,e_2) \ , \quad e_\ell \equiv max(e_1,e_2)
\label{emaxmin}
\eeq
Clearly,
\beq 
e_1+e_2=p
\label{eep}
\eeq
We denote this family of open ($o$) polygon chain graphs as 
$G_{e_1,e_2,e_g,m;o}$.  Illustrations of these graphs are given in Fig. 
(\ref{pgchain}). 
Since it is immaterial whether one labels the upper and lower routes
between the left and right-hand points on each polygon where it connects to the
rest of the chain as $e_1$ and $e_2$ or in the opposite order, as $e_2$ and
$e_1$, 
\beq
G_{e_1,e_2,e_g,m;o} = G_{e_2,e_1,e_g,m;o}
\label{gsymopen}
\eeq

The number of vertices is
\beq
v(G_{e_1,e_2,e_g,m;o})=(p+e_g-1)m+1
\label{nopen}
\eeq
and the number of edges or bonds is 
\beq
e(G_{e_1,e_2,e_g,m;o})=(p+e_g)m
\label{eopen}
\eeq
The chromatic number $\chi=2$ if $p$ is even and $\chi=3$ if $p$ is odd.  The
chromatic polynomial is easily calculated to be 
\beq
P(G_{e_1,e_2,e_g,m;o},q)=q (a_1)^m
\label{pgopen}
\eeq
where
\beq
a_1 = (q-1)^{e_g+1}D_p
\label{a1}
\eeq
Evidently, the chromatic polynomial has the form of (\ref{pgen}) with 
$c_0=0$, $N_a=1$, $c_1=q(a_1)^{-t_1}$ and $t_1=t$, where 
\beq
t = \frac{1}{p+e_g-1}
\label{t}
\eeq
Note that this chromatic polynomial does not depend on $e_1$ or $e_2$
individually, but instead only on their sum $p$, i.e., the number of sides 
of each polygon\footnote{
One can consider a generalization of this family in
which the quantities $e_1$ and $e_2$ can be different for each polygon.  The
chromatic polynomial and $W$ function for this much larger family of graphs are
again given by (\ref{pgopen}) with (\ref{a1}) and by (\ref{wopen}),
respectively.}.

Let us next consider the limit of an infinitely long chain, i.e., 
$m \to \infty$.  Formally, we denote this limit as 
\beq
L_m: \quad m \to \infty \quad {\rm with} \quad e_1 \ , e_2 \ , e_g \quad 
{\rm fixed}
\label{minf}
\eeq
We find 
\beq
W([\lim_{m \to \infty}G_{e_1,e_2,e_g,m;o}],q) = 
[(q-1)^{e_g+1}D_p]^{\frac{1}{p+e_g-1}}
\label{wopen}
\eeq
The nonanalyticities of this $W$ function consist of discrete branch point
singularities at its zeros, including the point $q=1$ and the 
$p-2$ zeros of $D_p$; there is no continuous locus of
nonanalyticities of $W$; i.e., ${\cal B} = \emptyset$.  If $p$ is odd, then 
$D_p$ contains a factor $(q-2)$, so that one of the zeros of $W$ is at $q=2$. 

Given the relation (\ref{nopen}), there are also 
other ways to take the number of vertices to infinity: 
\beq
L_{e_g}: \quad e_g \to \infty \quad {\rm with} \quad e_1 \ , e_2 \ , m 
\quad {\rm fixed}
\label{ginf}
\eeq
\beq
L_{e_1}: \quad e_1 \to \infty \quad {\rm with} \quad e_2 \ , e_g \ , m
\quad {\rm fixed}
\label{e1inf}
\eeq
\beq
L_{e_2}: \quad e_2 \to \infty \quad {\rm with} \quad e_1 \ , e_g \ , m 
\quad {\rm fixed}
\label{e2inf}
\eeq
and 
\beq
L_p: \quad p \to \infty   \quad {\rm with} \quad e_2-e_1 \ , e_g \ , m 
\quad {\rm fixed}
\label{pinf}
\eeq
For the $L_{e_g}$ limit, we find 
\beq
W([\lim_{e_g \to \infty}G_{e_1,e_2,e_g,m;o}],q) = q-1
\label{wopeneglim}
\eeq
This is analytic for all $q$ so that, in particular, ${\cal B}=\emptyset$. 
The $W$ function in eq. (\ref{wopeneglim}) is the same as the $W$ function 
for linear and, more generally tree, graphs.
This is understandable, since as $e_g \to \infty$, the vertices on the linear 
connecting segments occupy a fraction approaching unity of the total number 
of vertices of the graph.  

Since the chromatic 
polynomial (\ref{pgopen}) only depends on $e_1$ and $e_2$ through the 
quantity $p$, it follows that for this family of open polygon chain
graphs, the limits $L_{e_1}$, $L_{e_2}$, and $L_p$ all yield the same result.
Further, from eq. (\ref{dk}), it follows that in these limits, $W$ is the same
as that for a circuit graph \cite{w}, viz., 
\beq
W([\lim_{p \to \infty}G_{e_1,e_2,e_g,m;o}],q) = q-1 \quad {\rm for} \quad 
|q-1| \ge 1
\label{wopenpr1}
\eeq
and\footnote{
For real $0 < q < q_c(\{G\})$, as well as other regions of the $q$
plane that cannot be reached by analytic continuation from the real interval $q
> q_c(\{G\})$, one can only determine the magnitude $|W(\{G\},q)|$
unambiguously \cite{w}.}
\beq
|W([\lim_{p \to \infty}G_{e_1,e_2,e_g,m;o}],q)| = 1 \quad {\rm for} \quad 
|q-1| < 1
\label{wopenpr2}
\eeq
For this limit, ${\cal B}$ consists of the circle $|q-1|=1$.  Again, these
results are easily understood, since as $p \to \infty$, the vertices on the 
polygonal circuit subgraphs occupy a fraction approaching unity of the total 
number of vertices. 

\section{Cyclic Polygon Chain Graphs}

We proceed to our main subject, families of graphs consisting of $m$ $p$-sided
polygons linked together to form closed circuits. The other parameters, $e_g$,
$e_1$, and $e_2$, are the same as for the open polygonal chains discussed
above.  Thus, a given cyclic polygonal chain graph is specified by the four
parameters $(e_1,e_2,e_g,m)$, so we denote it as $G_{e_1,e_2,e_g,m}$, where the
cyclic property is implicitly understood.  For the same reason as given above
for the open chain, we have 
\beq
G_{e_1,e_2,e_g,m} = G_{e_2,e_1,e_g,m}
\label{gsym}
\eeq
The number of vertices in the graph
$G_{e_1,e_2,e_g,m}$ is 
\beq 
n = v(G_{e_1,e_2,e_g,m}) = (p+e_g-1)m
\label{ntot}
\eeq 
and the number of edges (bonds) is the same as for the open chain,
$e(G_{e_1,e_2,e_g,m})=(p+e_g)m$.  If $p$ is odd, then the chromatic number
$\chi$ is 3.  If $p$ is even, then $\chi=2$ or 3 depending on the values of
$e_1$, $e_2$, $e_g$, and $m$.  In Fig. (\ref{pgchain}) we show illustrative
examples of families of cyclic polygon chain graphs.  From (\ref{gsym}), it
follows that 
\beq 
P(G_{e_1,e_2,e_g,m},q) = P(G_{e_2,e_1,e_g,m},q)
\label{ppcsym}
\eeq

Using the deletion-contraction theorem to get recursion relations which we then
solve, we calculate the chromatic polynomial
\beq
P(G_{e_1,e_2,e_g,m},q) = (a_1)^m + (q-1)(a_2)^m
\label{ppc}
\eeq
where the term $a_1$ is the same as the term appearing in the chromatic
polynomial for the open polygonal chain, (\ref{a1}), and 
\beqs
a_2 & = & (-1)^{p+e_g}q^{-1}\Bigl [ q-2  + (1-q)^{e_1} + (1-q)^{e_2} \Bigr ] \\
& = &  (-1)^{p+e_g}\biggl [ 1-p - \sum_{s=2}^{e_1}{e_1 \choose s}(-q)^{s-1} 
- \sum_{s=2}^{e_2}{e_2 \choose s}(-q)^{s-1} \biggr ]
\label{a2}
\eeqs
The chromatic polynomial (\ref{ppc}) has the form of (\ref{pgen}) with 
$c_0=0$, $N_a=2$, $c_1=1$, $c_2=(q-1)$, and $t_1=t_2=t$, where $t$ was given in
eq. (\ref{t}).  Note that, owing to the second term, $a_2$, the chromatic
polynomial (\ref{ppc}) depends on the values of $e_1$ and $e_2$ individually,
in contrast to the result (\ref{pgopen}) for the open polygon chain, which only
depends on $e_1$ and $e_2$ through their sum, $p$. 

If the smaller distance $e_s = 1$, then 
the chromatic polynomial (\ref{ppc}) reduces to the factorized form 
\beq
P(G_{e_s=1,e_\ell,e_g,m},q) = (D_p)^mP(C_{n_c},q) = q(q-1)(D_p)^mD_{n_c}
\label{pfac}
\eeq
where
\beq
n_c=m(1+e_g)
\label{nc}
\eeq 
Having calculated the chromatic polynomial (\ref{ppc}), we can next obtain
the $W$ function and its continuous locus of nonanalyticities ${\cal B}$ for
the various limits (\ref{minf}) and (\ref{ginf})-(\ref{pinf}).  This locus is
determined as the solution of the degeneracy equation of (nonzero) leading
terms 
\beq
|a_1|=|a_2|
\label{degeneq}
\eeq
It will be useful to reexpress this in the variable $a=q-1$ of eq. (\ref{a}).
Multiplying both sides of (\ref{degeneq}) by $|a+1|$, we obtain
\beq
|a^{e_g}(a^p+(-1)^pa)|=|a-1+(-a)^{e_1}+(-a)^{e_2}|
\label{adegeneq}
\eeq
(Since the value $q=0$, i.e., $a+1=0$, is already a solution of 
(\ref{degeneq}), multiplying both sides of this equation by $|a+1|$ to get 
(\ref{adegeneq}) does not introduce a distinct spurious zero.) 

We first
discuss the limits that yield the simplest results.  For the $L_{e_g}$ limit,
${\cal B}$ consists of the unit circle $|q-1|=1$ and divides the $q$ plane into
two regions, the exterior and interior of this circle, which regions are
denoted as $R_1$ and $R_2$, respectively.  If $|q-1| > 1$, then $a_1$ is the
dominant term and 
\beq 
W([\lim_{e_g \to \infty}G_{e_1,e_2,e_g,m}],q) = q-1
\quad {\rm for} \quad |q-1| > 1
\label{wplimr1}
\eeq
If $|q-1| < 1$, then $a_2$ is dominant and 
\beq
|W([\lim_{e_g \to \infty}G_{e_1,e_2,e_g,m}],q)| = 1 \quad {\rm for} 
\quad |q-1| < 1
\label{wplimr2}
\eeq
That is, in the $L_g$ limit, the $W$ function is the same as for the
infinite-vertex limit of the circuit graph.  This is expected, since as $e_g$ 
increases, more and more of the vertices of the graph lie on a circuit graph
where, e.g., the route of the circuit follows either the $e_1$ or $e_2$ bonds
of a given polygon, then traverses the $e_g$ connecting bonds to the next
polygon, and so forth around the cyclic chain.  
As a result of the symmetry (\ref{ppcsym}), the $L_{e_1}$, $L_{e_2}$, and 
$L_p$ limits are equivalent; in this case, we find that for $|q-1| > 1$, the
$a_1$ term is dominant and $W$ is again given by the right-hand side of 
(\ref{wplimr1}) while for $|q-1| < 1$, neither term dominates over the other,
and $|W|=1$.  The results for $W$ are thus the same as for the $L_{e_g}$
limit, and the reason is similar: as $p \to \infty$, the vertices 
located on the polygons occupy a fraction approaching unity of all of the 
vertices. 

The most interesting and complicated results are for the $L_m$ limit. We have
proved a number of general properties of ${\cal B}$.  We show calculations of
${\cal B}$ and comparison with chromatic zeros for long finite chains in
Figs. 2-5 (and, in addition, Figs. 6-9 in the Appendix).  Because of the
factorization (\ref{pfac}), the families with $e_s=1$ yield very simple results
(see below) so that in the figures we concentrate on the cases with $e_s \ge
2$: $(e_1,e_2)=(2,2)$, (2,3), (2,4), and (3,3) and, for each pair $(e_1,e_2)$,
the values $0 \le e_g \le 3$.  The general properties of ${\cal B}$ are 

\begin{itemize}

\item

(B1) ${\cal B}$ is compact.

\item 

(B2) ${\cal B}$ passes through $q=0$. 

\item 

(B3) ${\cal B}$ encloses regions in the $q$ plane. 

\item

(B4) if $e_s = 1$, then ${\cal B}$ is the unit circle $|q-1|=1$ independent
of the values of $e_\ell$ and $e_g$; thus, in this case, $q_c=2$.  The
chromatic zeros of $G_{e_1,e_2,e_g,m}$ lie exactly on this locus, independently
of the values of $e_\ell$, $e_g$, and $m$.

\item 

(B5) if $p$ is even, then $q_c=2$.

\item 

(B6) if $p$ is odd and $e_s \ge 2$, then $q_c < 2$ and for fixed $(e_1,e_2)$, 
$q_c$ increases monotonically as $e_g$ increases, approaching 2 from below
as $e_g \to \infty$

\end{itemize}

To prove (B1), we re-express eq. (\ref{degeneq}) in terms of the variable 
\beq
y= \frac{1}{a} = \frac{1}{q-1}
\label{ya}
\eeq
Then eq. (\ref{adegeneq}) becomes 
\beq
|1+(-1)^py^{p-1}| = |y^{e_g+e_s}(y^{e_\ell-1}-y^{e_\ell}+(-1)^{e_\ell}+
(-1)^{e_s}y^{e_\ell-e_s})|
\label{degeneqy}
\eeq
where we have divided (\ref{degeneq}) by $|a^{e_g+p}|$, which factor is
nonzero for $q \ne 1$ and thus, in particular, for the large--$|q|$ (small
$|y|$) region of interest here.  The necessary and sufficient condition that
${\cal B}$ is noncompact, i.e.  unbounded is that the equation
(\ref{degeneqy}) has a solution for $y=0$.  Clearly, it does not have such a
solution, which proves property (B1).

For (B2), we use the property that 
\beq
D_p(q=0)=(-1)^p(p-1)
\label{dpq0}
\eeq
so that $|a_1|=|p-1|$ at $q=0$.  From the second equivalent form of $a_2$ in 
eq. (\ref{a2}), it follows that $|a_2|=|p-1|$ at $q=0$ also, so that the point
$q=0$ is a solution to eq. (\ref{degeneq}) and hence is on the locus ${\cal
B}$, which proves (B2).  

Property (B3) can be proved by explicit solution of (\ref{degeneq}).  One also
observes from the results presented in Figs. 2-9 that in cases where ${\cal B}$
does not contain any multiple points\footnote{ In the technical terminology of
algebraic geometry \cite{alg}, a multiple point on an algebraic curve is a
point where several branches of the curve intersect.  If all $n_i$ of the
branches have different tangents, the multiple point is said to have index
$n_i$.}, the number of regions, $N_{reg.}$ and the number of connected
components on ${\cal B}$ satisfy the relation $N_{reg.}=N_{comp.}+1$.  As was
the case in our earlier work \cite{wa2}, one can get a general upper bound on
the number $N_{comp.}$ using the Harnack theorem \cite{alg} from algebraic
geometry, but it is not very restrictive.  To do this, we write the equation
(\ref{degeneq}) whose solution set is ${\cal B}$ out in terms of the real and
imaginary components of $q$ or, more conveniently, of $a=q-1$. This yields a
polynomial equation of general degree $d=2(p+e_g)$ in these components.  In the
cases without singular (multiple) points on ${\cal B}$, the Harnack theorem
then gives the upper bound $N_{comp.} \le h+1$, where $h$, the
genus of the algebraic curve comprising ${\cal B}$, is $h=(d-1)(d-2)/2$
\cite{alg}.  Thus, for example, for the lowest case $(e_1,e_2,e_g)=(2,2,1)$,
the Harnack theorem gives the upper bound $N_{comp.} \le 37$.  Clearly, this is
a very weak upper bound compared with the actual result $N_{comp.}=2$. The
upper bound becomes even higher for larger values of $e_g$, and hence weaker,
since the result remains $N_{comp.}=2$ for all families of the form
$(e_1,e_2,e_g)=(2,2,e_g)$.  Similar remarks apply for the Harnack bound applied
to other cases $(e_1,e_2,e_g)$.

To prove (B4), we observe that for families where $e_s=1$, because of the
factorization (\ref{pfac}), for any values of $e_\ell$ and $e_g$, the locus
${\cal B}$ is the unit circle $|q-1|=1$. The result on chromatic zeros follows
immediately from (\ref{pfac}) and the fact that the zeros of the chromatic
polynomial of the circuit graph lie exactly on the circle $|q-1|=1$ \cite{wc}.

To prove property (B5), we first observe that since $p$ is even, the left-hand
side of eq. (\ref{adegeneq}) has the value 2 for $q=2$, i.e., for $a=1$.  The
condition that $p$ is even means that either both $e_1$ and $e_2$ are even or
both $e_1$ and $e_2$ are odd.  In both of these cases, the right-hand side of
eq. (\ref{adegeneq}) is also equal to 2 for $q=2$, so that $q=2$ is a solution
of this equation. Furthermore one easily checks that (\ref{adegeneq}) has no
solution for larger real $q$.  This yields (B5).

For property (B6) pertaining to odd $p$, we recall that the case $e_s = 1$ is
already covered by (B4); for $e_s \ge 2$, we first note that since $p$ is odd,
one of $(e_1,e_2)$ is odd and the other is even; these are denoted $e_{even}$ 
and $e_{odd}$.  We next observe that for odd $p=2k+1$, $D_p$ contains a factor
$(q-2)$ times a polynomial:
\beq
D_{2k+1} = (q-2)F_{2k+1}
\label{dpodd}
\eeq
where
\beq
F_{2k+1} = k \quad {\rm at} \quad q=2
\label{fq2}
\eeq
Now let us consider the limit $q \to 2$.  From eqs. (\ref{dpodd}) and 
(\ref{fq2}), we have 
\beq
a_1 \to \frac{(q-2)(p-1)}{2} \quad {\rm as} \quad q \to 2
\label{a1limit}
\eeq
Furthermore, 
\beq
a_2 \to \frac{1}{2}(-1)^{e_g+1}(q-2)(1+e_{even}-e_{odd})  \quad {\rm as} 
\quad q \to 2
\label{a2limit}
\eeq
Hence for the chromatic polynomial (\ref{ppc}), 
\beq
P(G_{e_{even},e_{odd},e_g,m},q \to 2) 
\to \Biggl [\frac{(q-2)(p-1)}{2}\Biggr ]^m
\Biggl [ 1 + \biggl ( \frac{(-1)^{e_g+1}(1+e_{even}-e_{odd}}{p-1} \biggr )^m 
\Biggr ]
\label{ppodd}
\eeq
Since 
\beq
\frac{|1+e_{even}-e_{odd}|}{p-1} < 1
\label{eeineq}
\eeq 
for the relevant case $e_s \ge 2$ considered here, 
it follows that, taking the $m \to \infty$ limit first, before the $q \to
2$ limit, the contribution of the second term in the square brackets, i.e., the
$a_2$ term, in eq. (\ref{ppodd}), relative to that of the first, i.e., $a_1$
term, goes to zero.  Recalling our order of limits (\ref{wnoncom}), it follows
that at $q=2$, the function $W$ is determined completely by the $a_1$ term.
Hence, although $a_1$ and $a_2$ both vanish at $q=2$, this point is not on the
locus ${\cal B}$ since $W$ is determined only by $a_1$ at this point.  
This shows that the condition of the degeneracy of terms $|a_1(q)|=|a_2(q)|$ 
is necessary in order that the point $q$ lie on the locus ${\cal B}$, but, 
as noted in connection with eq. (\ref{degeneq}), a
further necessary (and sufficient) condition is that this equality in
magnitudes holds where $a_1$ and $a_2$ are both nonzero leading terms.
Since $W$ is determined by $a_1$ at $q=2$, the same as in region
$R_1$ (see below), this shows that $q_c < 2$, which was to be proved.

The other property in (B6) follows in a similar manner from the solution to
eq. (\ref{adegeneq}).  The fact that $q_c \to 2$, i.e., $a_c \to 1$, where
$a_c=q_c-1$, is clear from (\ref{adegeneq}) since if $a > 1$ ($a < 1$), the
left-hand side diverges (vanishes) as $e_g \to \infty$, precluding equality 
with the right-hand side.

\begin{figure}
\centering
\leavevmode
\epsfxsize=3.5in
\begin{center}
\leavevmode
\epsffile{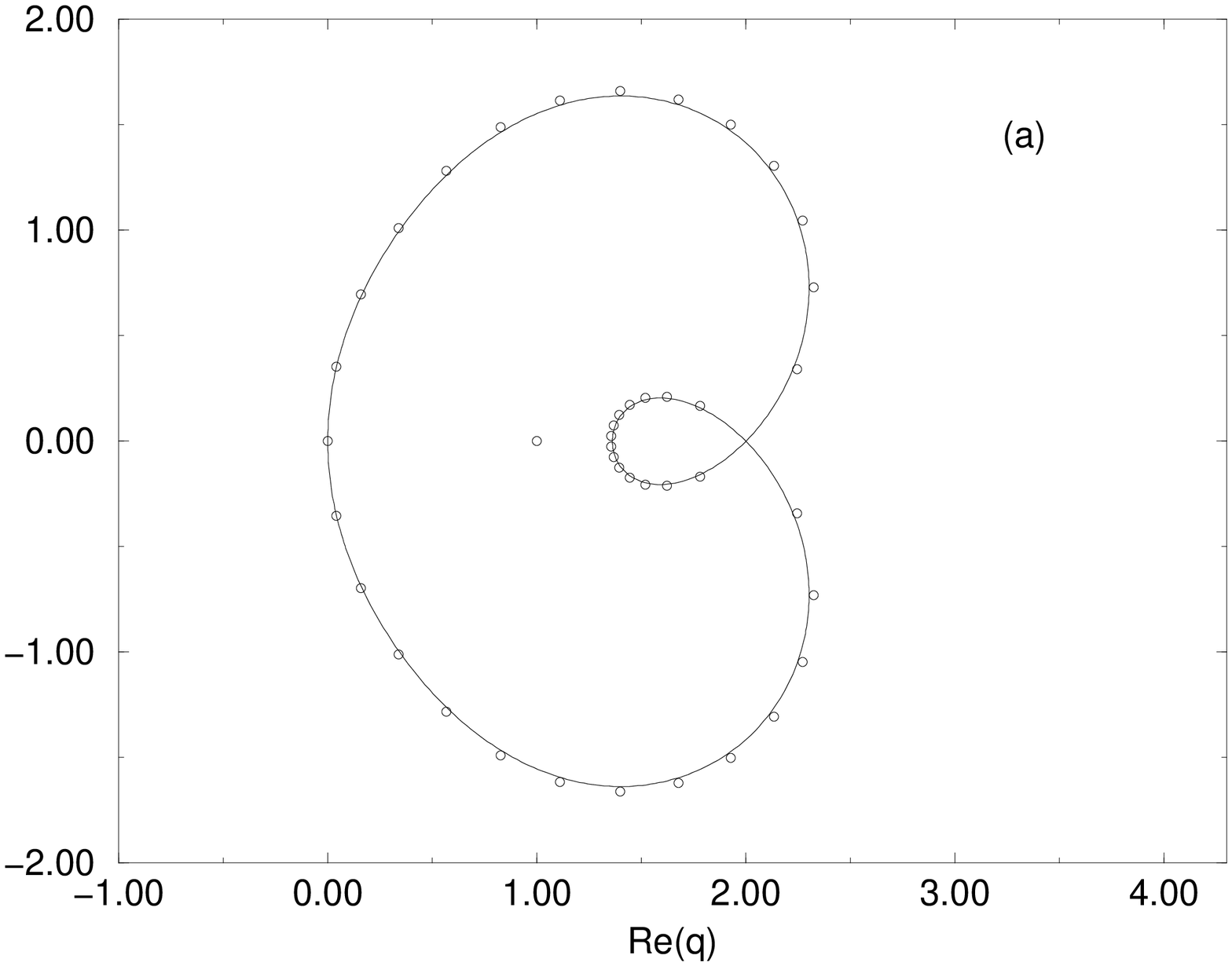}
\end{center}
\vspace{-4cm}
\begin{center}
\leavevmode
\epsfxsize=3.5in
\epsffile{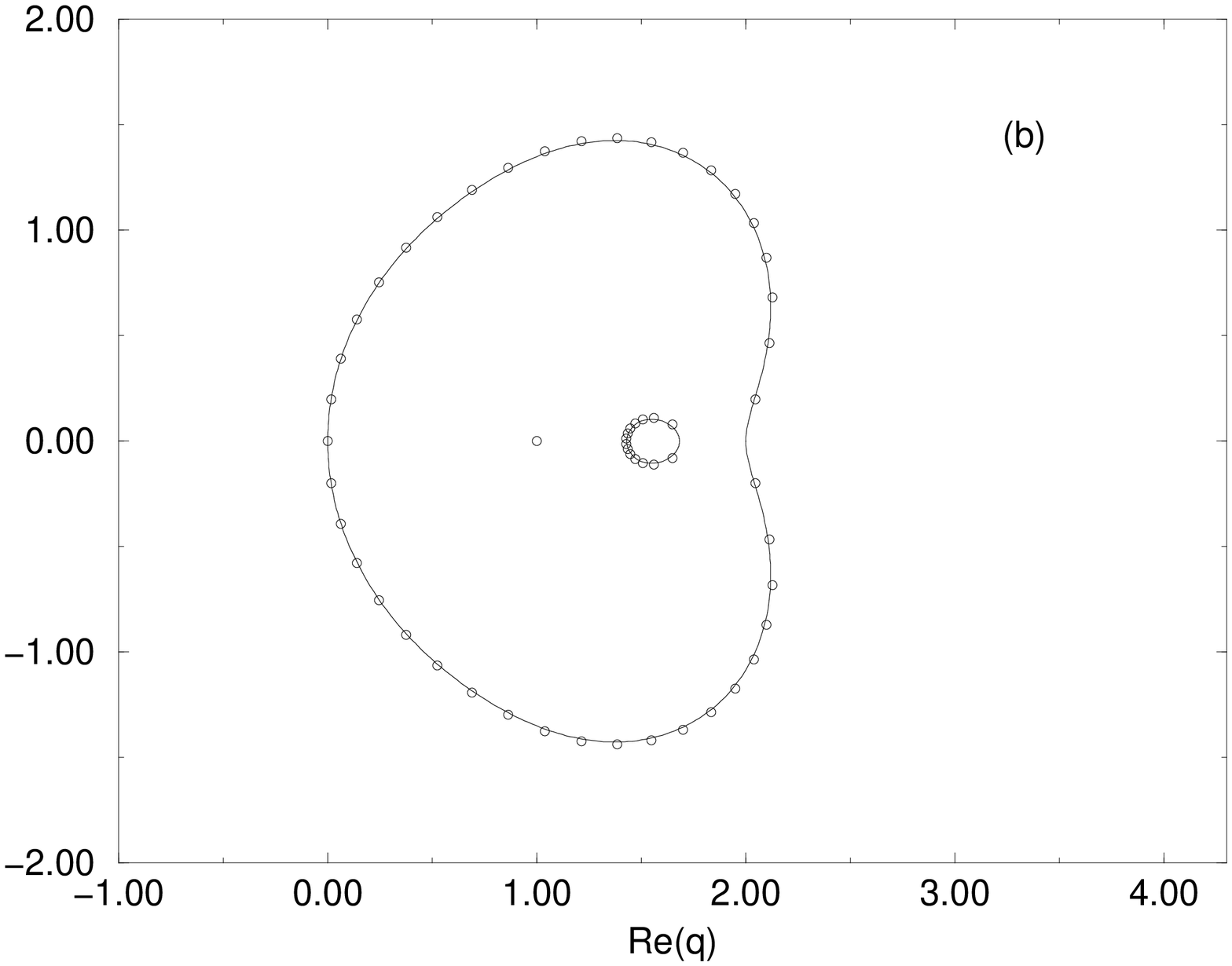}
\end{center}
\vspace{-2cm}
\caption{\footnotesize{Boundary ${\cal B}$ in the $q$ plane for $W$ function 
for $\lim_{m \to \infty} G_{e_1,e_2,e_g,m}$ with $(e_1,e_2,e_g)=$ 
(a) (2,2,0), (b) (2,2,1). Chromatic zeros for $m=14$ (i.e., $n=42$ and 
$n=56$ for (a) and (b)) are shown for comparison.}}
\label{nec22g01}
\end{figure}

We observe several other interesting features of the locus ${\cal B}$.  For
$e_g=0$ and $e_s \ge 2$ and for certain values of $(e_1,e_2)$, this locus, as
an algebraic curve, may contain one or more multiple point(s), which are of
index 2, i.e., they involve a crossing of four curves forming two branches of
${\cal B}$.  A particularly simple set of families for which this occurs is the
set where $e_1=e_2=p/2 \equiv e_{12}$.  By an analysis of eq. (\ref{adegeneq}),
we find that for this set there are 
\beq 
N_{m.p.} = e_{12}-1
\label{pevendiagmp}
\eeq
multiple points, which occur at 
\beq
q_j = 1 + e^{i\theta_j}
\label{qjmp}
\eeq
where
\beq
\theta_j = \pm \frac{2 j \pi}{e_{12}}, \quad j = 0, 1, ..., \frac{e_{12}-2}{2} 
\quad {\rm for} \quad e_{12} \ \ {\rm even} 
\label{qje12even}
\eeq
\beq
\theta_j = \pm \frac{(2j+1)\pi}{e_{12}}, \quad j = 0, 1, ..., 
\frac{e_{12}-3}{2} \quad {\rm for} \quad e_{12} \ \ {\rm odd}
\label{qje12odd}
\eeq 
For example, in Fig. (\ref{nec22g01})(a) for $(e_1,e_2,e_g)=(2,2,0)$ one sees
that there is one such multiple point, located at $q=q_c=2$, while in
Fig. (\ref{nec33g01})(a) for $(e_1,e_2,e_g)=(3,3,0)$ there are 2
complex-conjugate multiple points, located at $q=1+e^{\pm i \pi/3}$.  Fig. 
(\ref{nec24g01}) illustrates another case with a multiple point, namely 
$(e_1,e_2,e_g)=(2,4,0)$, for which the multiple point is at $q=q_c=2$.  In
Figs. 2-3 and 6-9 one sees that the multiple points that are present for 
$e_g=0$ disappear when $e_g \ge 1$.  Figs. 4-5 illustrate a family with 
$p$ odd, viz., $(e_1,e_2)=(2,3)$ for which ${\cal B}$ has no multiple points. 

\begin{figure}
\centering
\leavevmode
\epsfxsize=3.5in
\begin{center}
\leavevmode
\epsffile{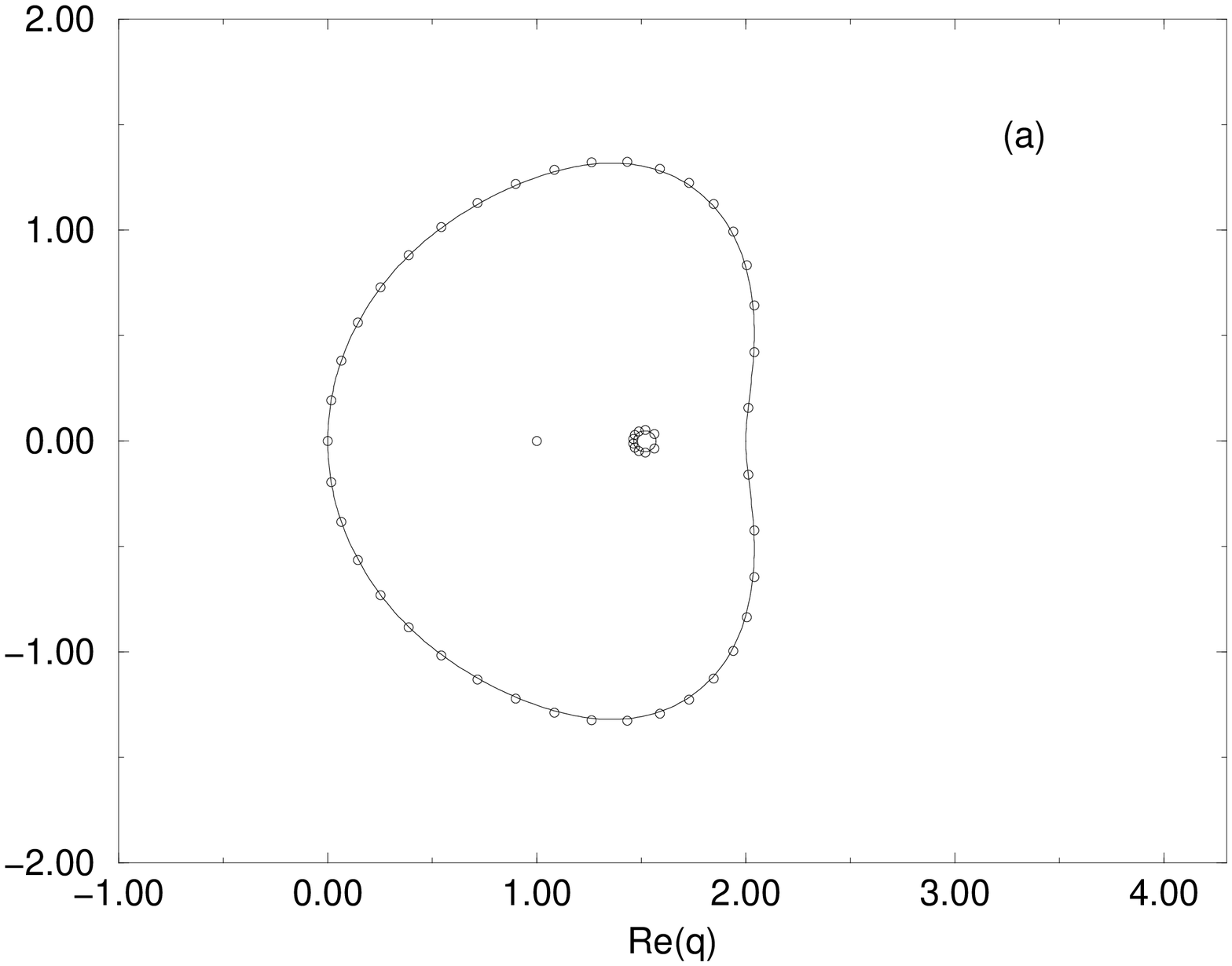}
\end{center}
\vspace{-4cm}
\begin{center}
\leavevmode
\epsfxsize=3.5in
\epsffile{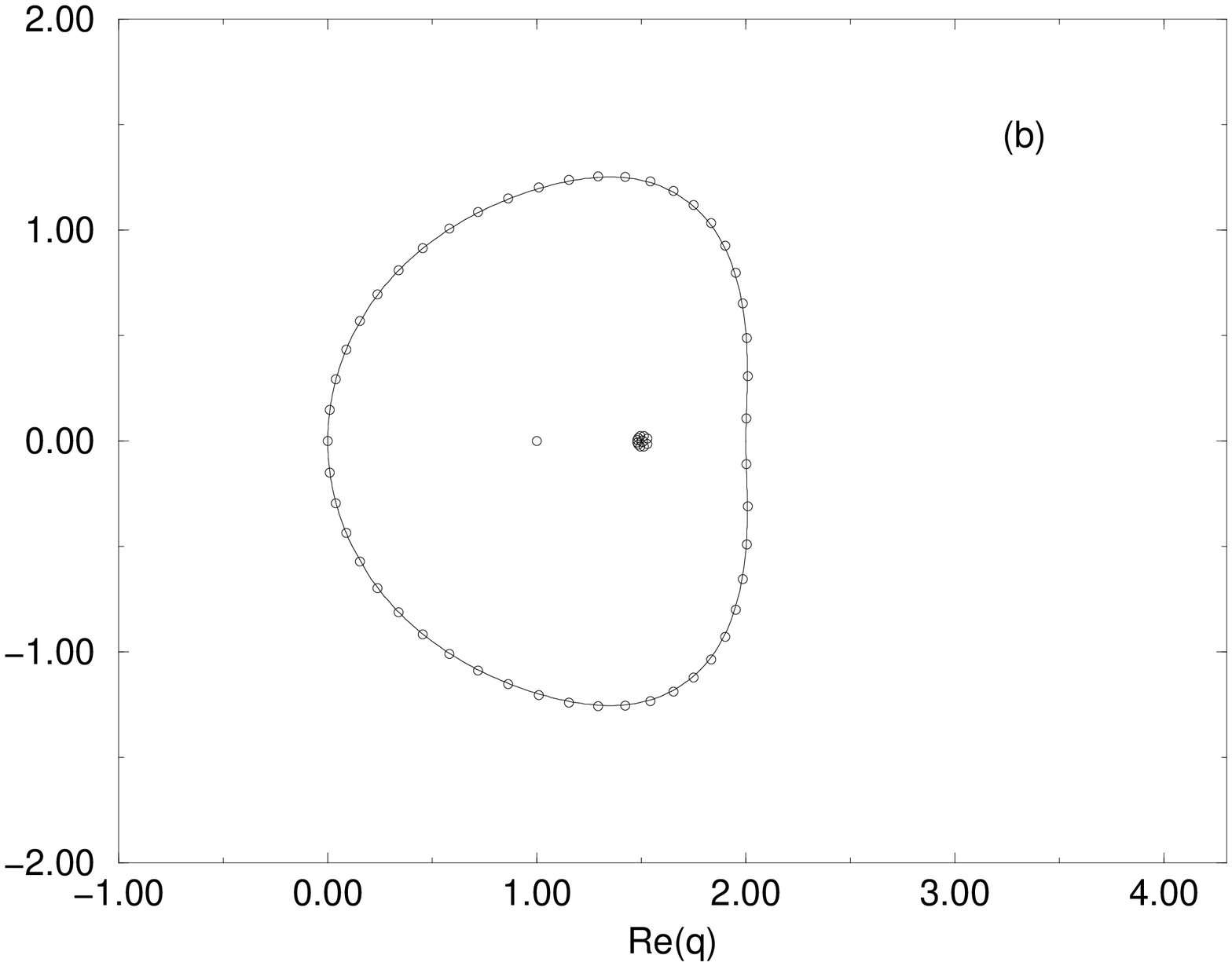}
\end{center}
\vspace{-2cm}
\caption{\footnotesize{Boundary ${\cal B}$ in the $q$ plane for $W$ function
for $\lim_{m \to \infty} G_{e_1,e_2,e_g,m}$ with $(e_1,e_2,e_g)=$
(a) (2,2,2), (b) (2,2,3). Chromatic zeros for $m=10$ (i.e., $n=50$ and
$n=60$ for (a) and (b)) are shown for comparison.}}
\label{nec22g23}
\end{figure}

In the cases where ${\cal B}$ contains multiple point(s) for $e_g=0$, which
disappear when $e_g \ge 1$, each such disappearance is accompanied by the
appearance of a new disconnected component of ${\cal B}$.  As $e_g$ increases,
the sizes of the regions bounded by these components of ${\cal B}$ decrease,
and as $e_g \to \infty$, the regions shrink to points.  For example, in
Figs. (\ref{nec22g01}) and (\ref{nec22g23}) one sees that for the case
$(2,2,e_g)$, as $e_g$ increases from $e_g=0$ to $e_g \ge 1$, the multiple point
at $q=q_c=2$ that was present for $e_g=0$ disappears and there appears a
disconnected component of ${\cal B}$ defining a new region, so that the number
$N_{comp.}$ of (separate, disconnected) components of ${\cal B}$ increases from
1 to 2.  As $e_g \to \infty$, this region shrinks to a point at $q=3/2$.
Similarly, in Figs. (\ref{nec33g01}) and (\ref{nec33g23}) one sees that for the
case $(e_1,e_2)=(3,3)$, as $e_g$ increases from 0 to $e_g \ge 1$, the
complex-conjugate multiple points that were present for $e_g=0$ at $q=1+e^{\pm
i \pi/3}$ disappear and $N_{comp.}$ increases from 1 to 3.  As $e_g \to
\infty$, the inner regions shrink to points at $q \simeq 3/2 \pm i/2$.

Another feature that one observes in the results shown in the figures is that
these loci ${\cal B}$ do not have any support for $Re(q) < 0$.  This is in
accord with our conjecture \cite{strip,pg} that global circuits are a necessary
but not sufficient condition for ${\cal B}$ to have support for $Re(q) < 0$. 

For $W([\lim_{m \to \infty}G_{e_1,e_2,e_g,m}],q)$, which henceforth will be
abbreviated simply as $W$, we have the following general results.  
In region $R_1$, 
\beq
W = (a_1)^{\frac{1}{p+e_g-1}} = 
[(q-1)^{e_g+1}D_p]^{\frac{1}{p+e_g-1}} \quad {\rm for} \quad q \in R_1 
\label{wr1}
\eeq
We note that the function on the 
right-hand side of eq. (\ref{wr1}) is the same as that for the open chain, 
(\ref{wopen}), although in the latter case, the result holds throughout the
full $q$ plane whereas for the cyclic chain, the result holds only in region
$R_1$.  One can approach the origin, $q=0$ from the left staying within the 
region $R_1$, and in this limit,
using the property (\ref{dpq0}), we have 
\beq
W=(p-1)^{\frac{1}{p+e_g-1}} \quad {\rm for} \quad q \to 0 \ , \quad q \in R_1
\label{wq0}
\eeq
Let us denote region $R_2$ as the one containing the point $q=1$; this region
is contiguous to $R_1$ at the origin, $q=0$.  In region $R_2$, 
$a_2$ is the dominant term, so that 
\beq
|W|=|a_2|^{\frac{1}{p+e_g-1}} \quad {\rm for} \quad q \in R_2
\label{wr2}
\eeq
In particular, for families of graphs with $e_s=1$, where the factorization 
(\ref{pfac}) holds, 
\beq
|W|=|D_p|^{\frac{1}{p+e_g-1}} \quad {\rm for} \quad e_s=1 \quad {\rm and} 
\quad q \in R_2
\label{wr2es1r2}
\eeq
For cases where there are additional regions $R_j$ enclosed within $R_2$, 
\beq
|W|= |a_1|^{\frac{1}{p+e_g-1}} \quad {\rm for} \quad q \in R_j \quad 
{\rm enclosed \ within} \quad R_2
\label{wrj}
\eeq
For $e_g=0$ these additional regions are contiguous with $R_1$ at multiple
points on ${\cal B}$.  

Concerning values of $W$ at special points, we observe that since $q_c \le 2$,
the expression (\ref{wr1}) always holds at $q=2$, so that, as a consequence of
the property (\ref{dpodd}), 
\beq
W(q=2)=0 \quad {\rm for} \quad {\rm odd} \quad p
\label{wpoddq2}
\eeq
Note that this does not follow just from the property that the chromatic 
polynomial $P$ vanishes at $q=2$ for odd $p$, but requires the stronger 
property that $P$ contains the factor $(q-2)^\nu$ with $\nu \propto n$ as 
$n \to \infty$. In contrast, for even $p=2k$, using the property
\beq
D_{2k} = 1 \quad {\rm for} \quad q=2
\label{dpevenq2}
\eeq
we have 
\beq
W(q=2)=1 \quad {\rm for} \quad {\rm even} \quad p
\label{wpevenq2}
\eeq
As one approaches $q=0$ from within region $R_2$, using (\ref{dpq0}), it
follows that 
\beq 
|W|=|p-1|^{\frac{1}{p+e_g-1}} \quad {\rm for} \quad q \to 0 \ , \quad q \in R_2
\label{wq0r2}
\eeq 
in accord with (\ref{wq0}) and the equality of the magnitude of $|W|$
across any point on ${\cal B}$.

\begin{figure}
\centering
\leavevmode
\epsfxsize=3.5in
\begin{center}
\leavevmode
\epsffile{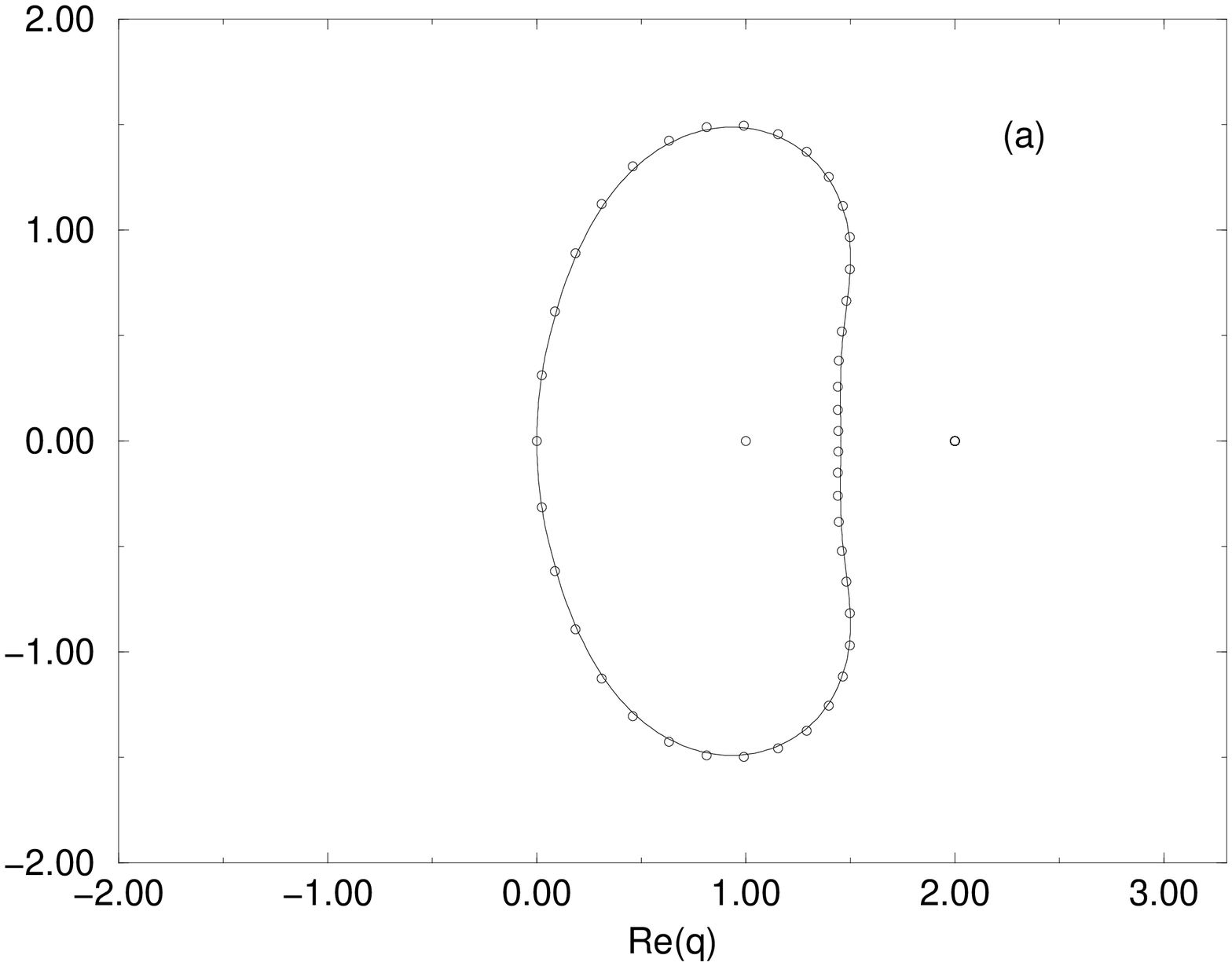}
\end{center}
\vspace{-4cm}
\begin{center}
\leavevmode
\epsfxsize=3.5in
\epsffile{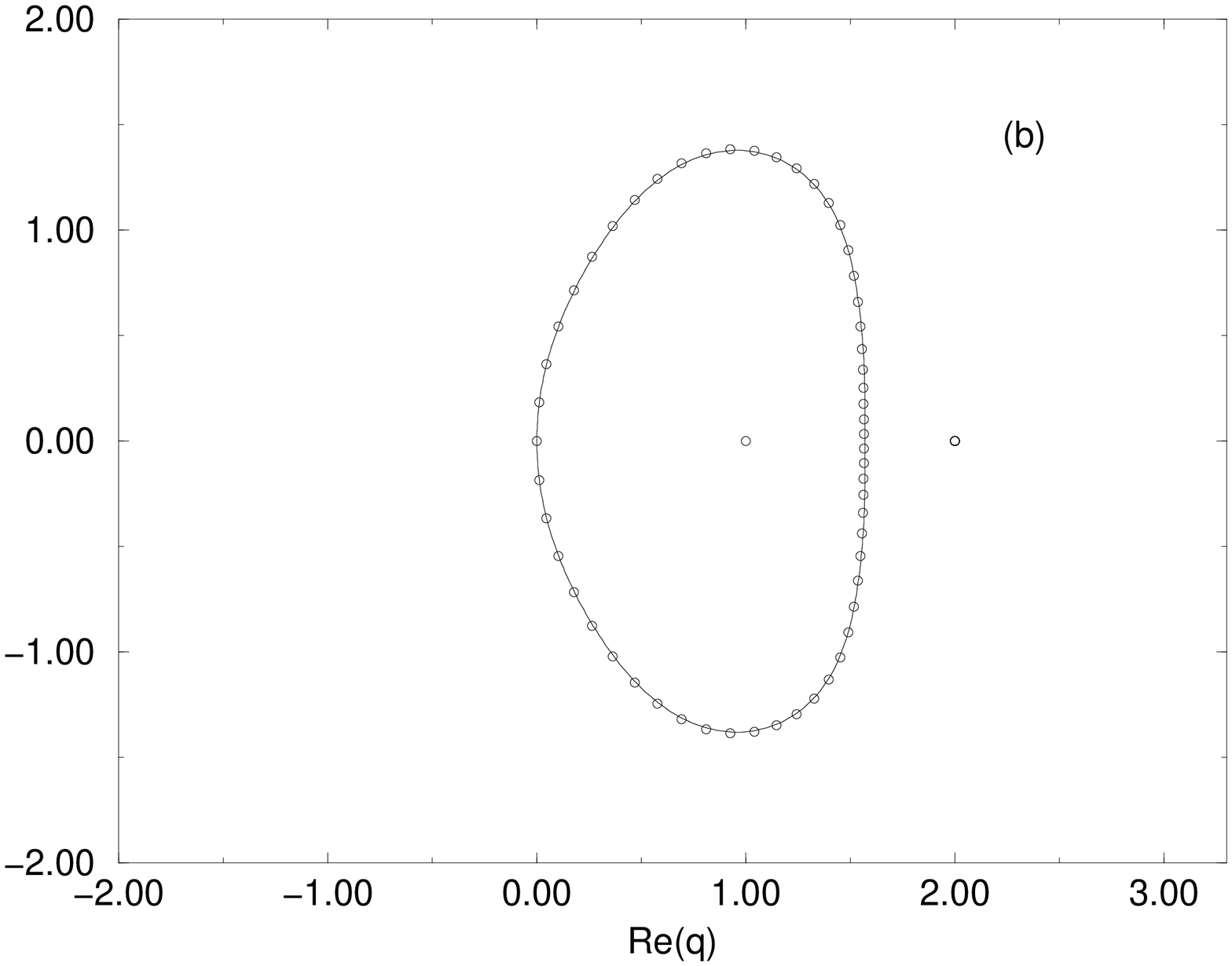}
\end{center}
\vspace{-2cm}
\caption{\footnotesize{Boundary ${\cal B}$ in the $q$ plane for $W$ function
for $\lim_{m \to \infty} G_{e_1,e_2,e_g,m}$ with $(e_1,e_2,e_g)=$
(a) (2,3,0), (b) (2,3,1). Chromatic zeros for $m=14$ (i.e., $n=56$ and
$n=70$ for (a) and (b)) are shown for comparison.}}
\label{nec23g01}
\end{figure}

\begin{figure}
\centering
\leavevmode
\epsfxsize=3.5in
\begin{center}
\leavevmode
\epsffile{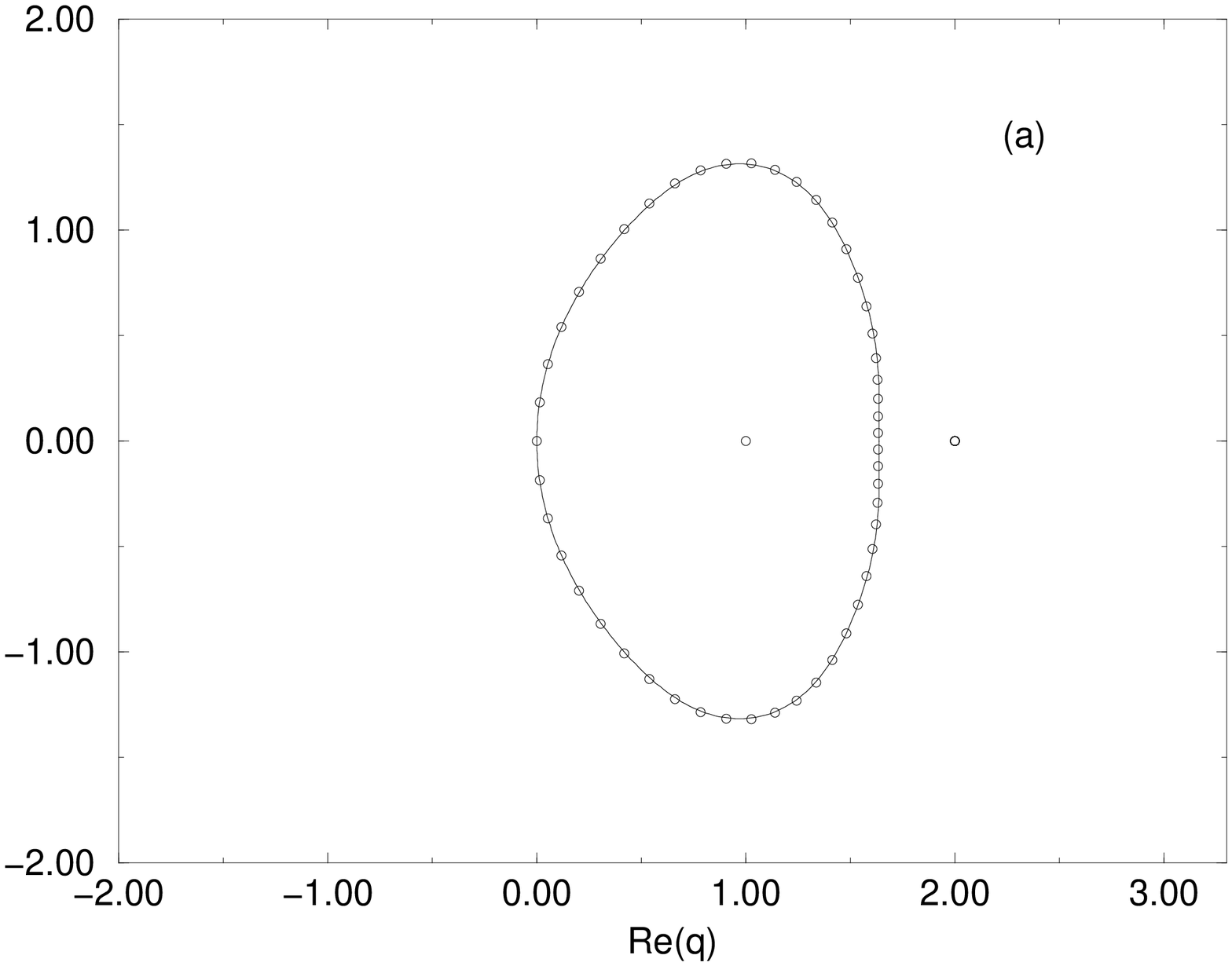}
\end{center}
\vspace{-4cm}
\begin{center}
\leavevmode
\epsfxsize=3.5in
\epsffile{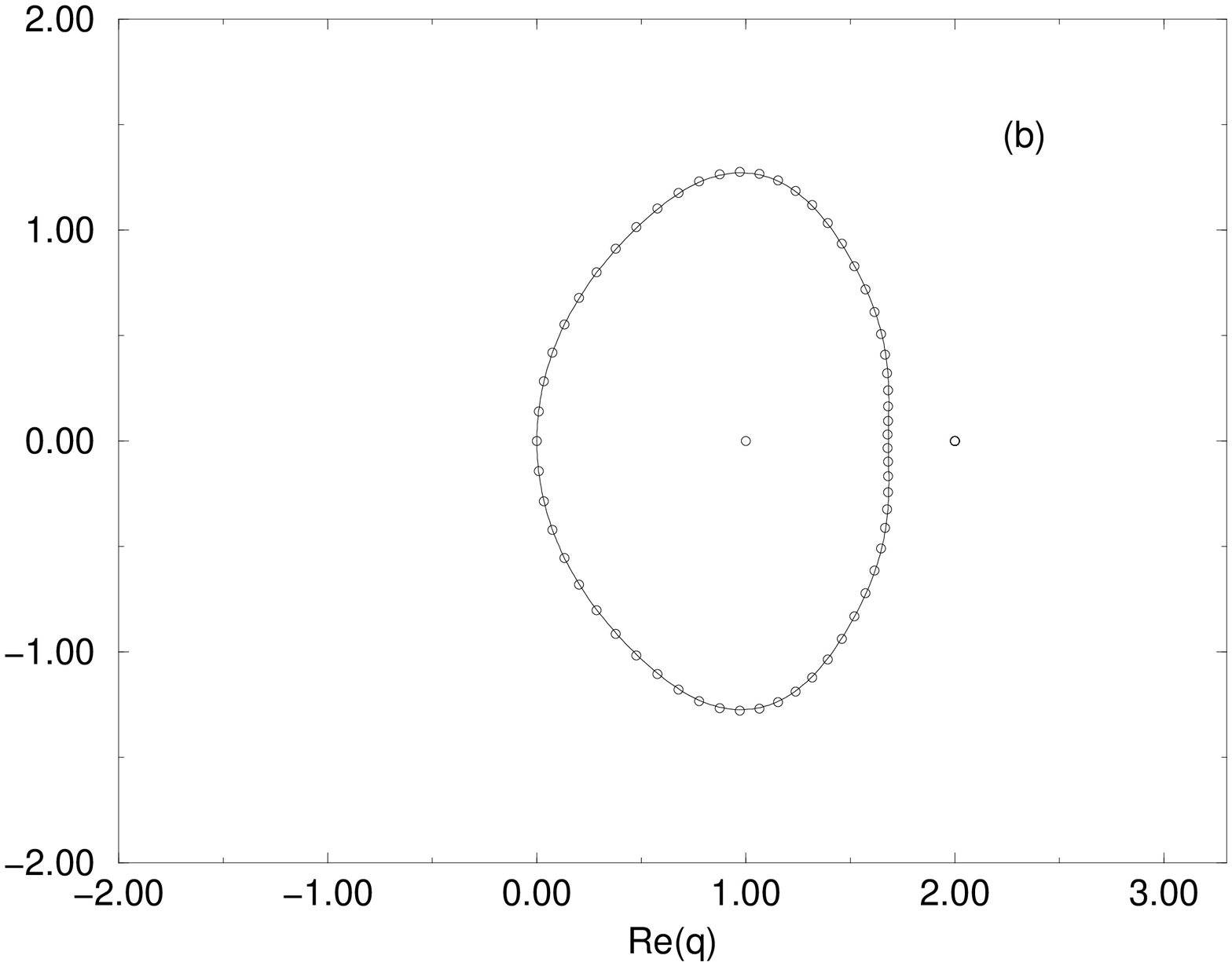}
\end{center}
\vspace{-2cm}
\caption{\footnotesize{Boundary ${\cal B}$ in the $q$ plane for $W$ function
for $\lim_{m \to \infty} G_{e_1,e_2,e_g,m}$ with $(e_1,e_2,e_g)=$
(a) (2,3,2), (b) (2,3,3). Chromatic zeros for $m=10$ (i.e., $n=60$ and
$n=70$ for (a) and (b)) are shown for comparison.}}
\label{nec23g23} 
\end{figure}

\begin{figure}
\centering
\leavevmode
\epsfxsize=3.5in
\begin{center}
\leavevmode
\epsffile{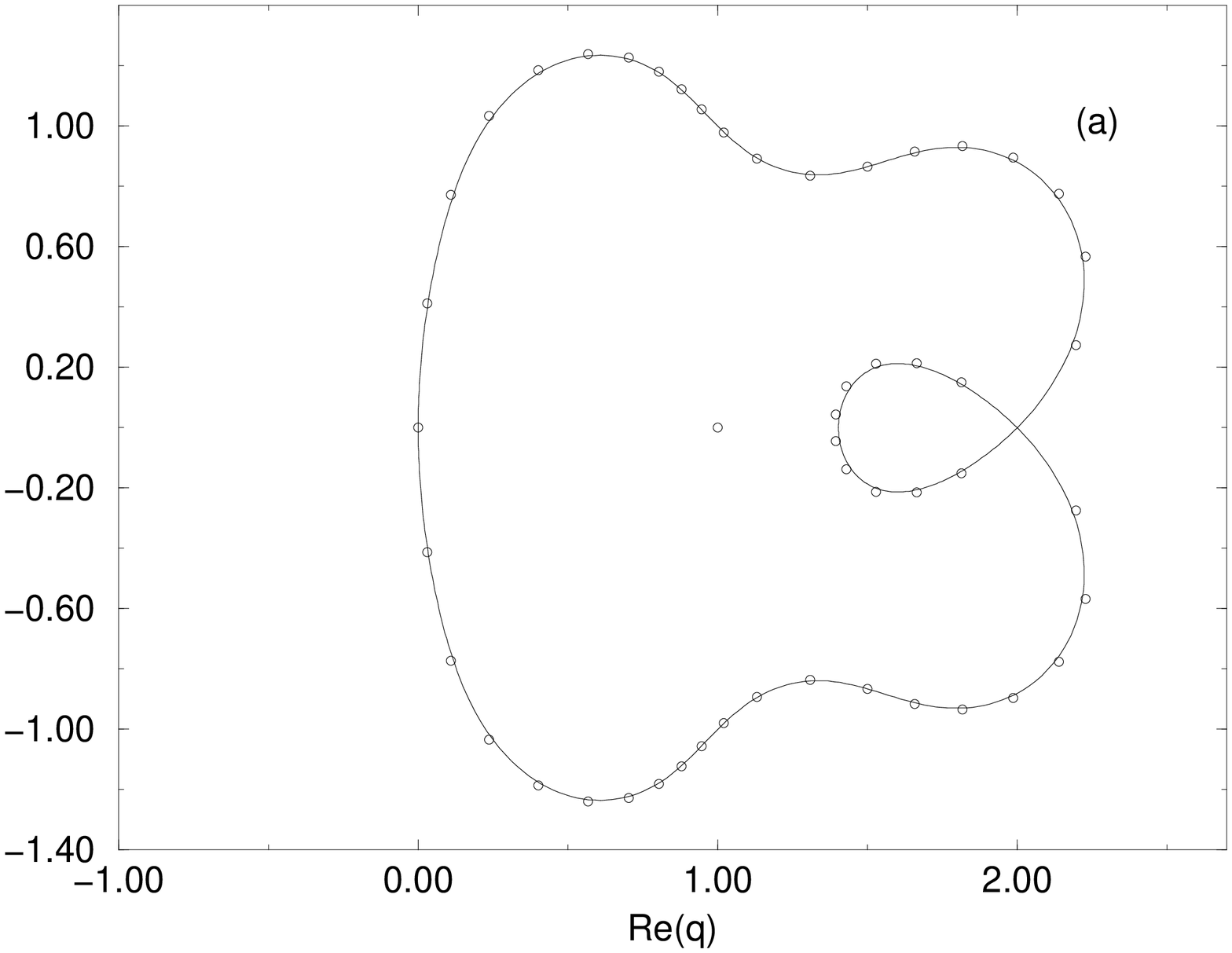}
\end{center}
\vspace{-4cm}
\begin{center}
\leavevmode
\epsfxsize=3.5in
\epsffile{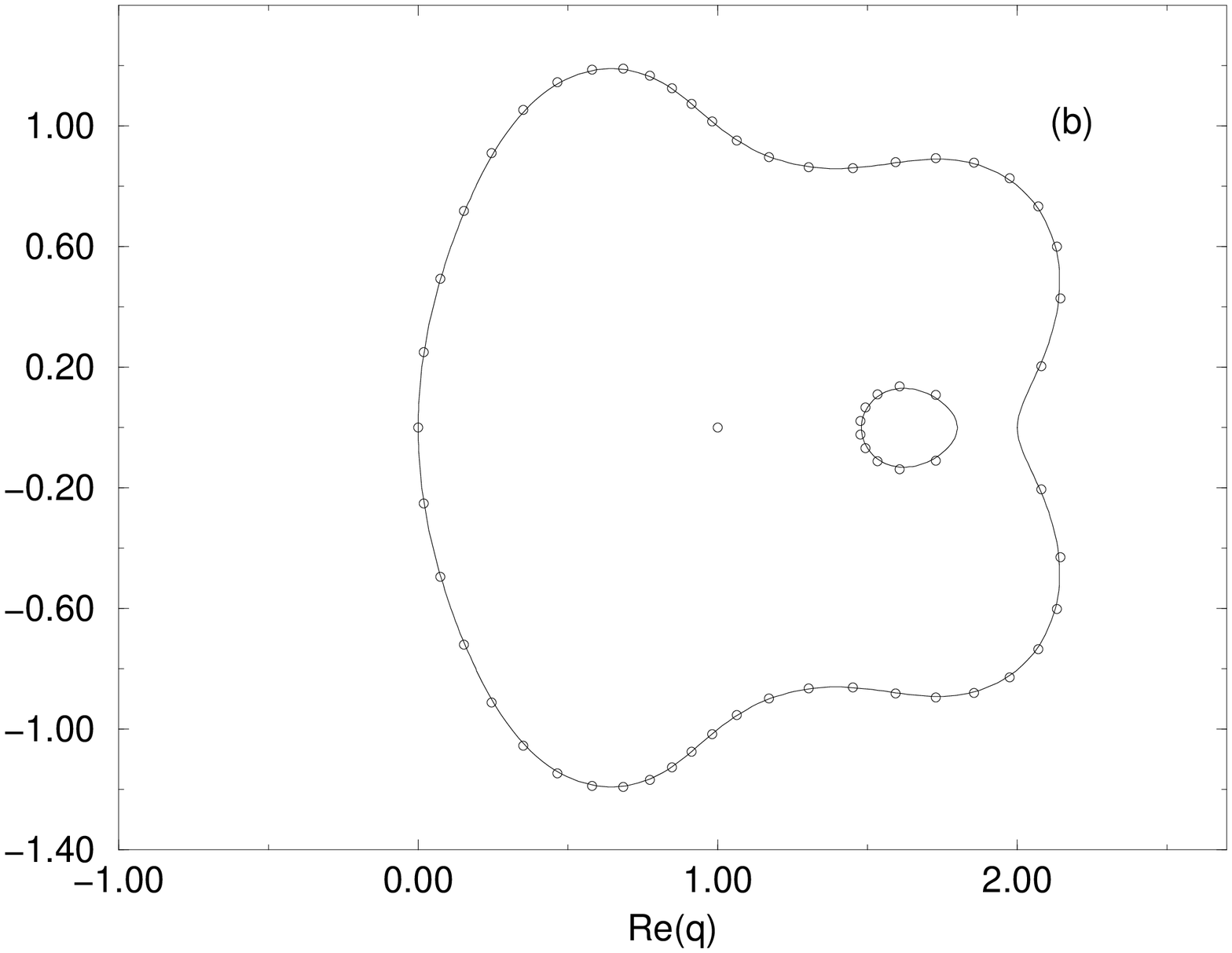}
\end{center}
\vspace{-2cm}
\caption{\footnotesize{Boundary ${\cal B}$ in the $q$ plane for $W$ function
for $\lim_{m \to \infty} G_{e_1,e_2,e_g,m}$ with $(e_1,e_2,e_g)=$
(a) (2,4,0), (b) (2,4,1). Chromatic zeros for $m=10$ (i.e., $n=50$ and
$n=60$ for (a) and (b)) are shown for comparison.}}
\label{nec24g01}
\end{figure}

\begin{figure}
\centering
\leavevmode
\epsfxsize=3.5in
\begin{center}
\leavevmode
\epsffile{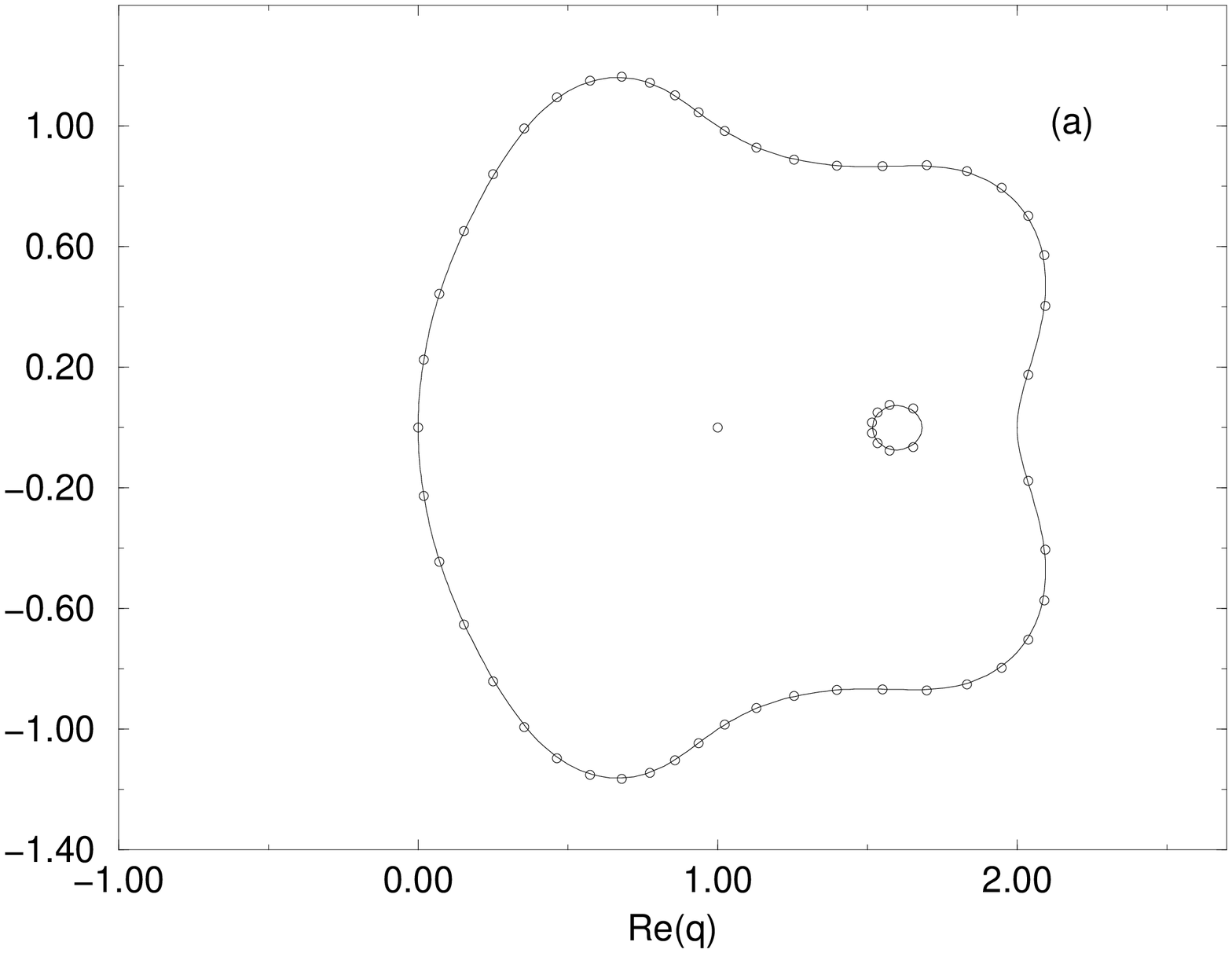}
\end{center}
\vspace{-4cm}
\begin{center}
\leavevmode
\epsfxsize=3.5in
\epsffile{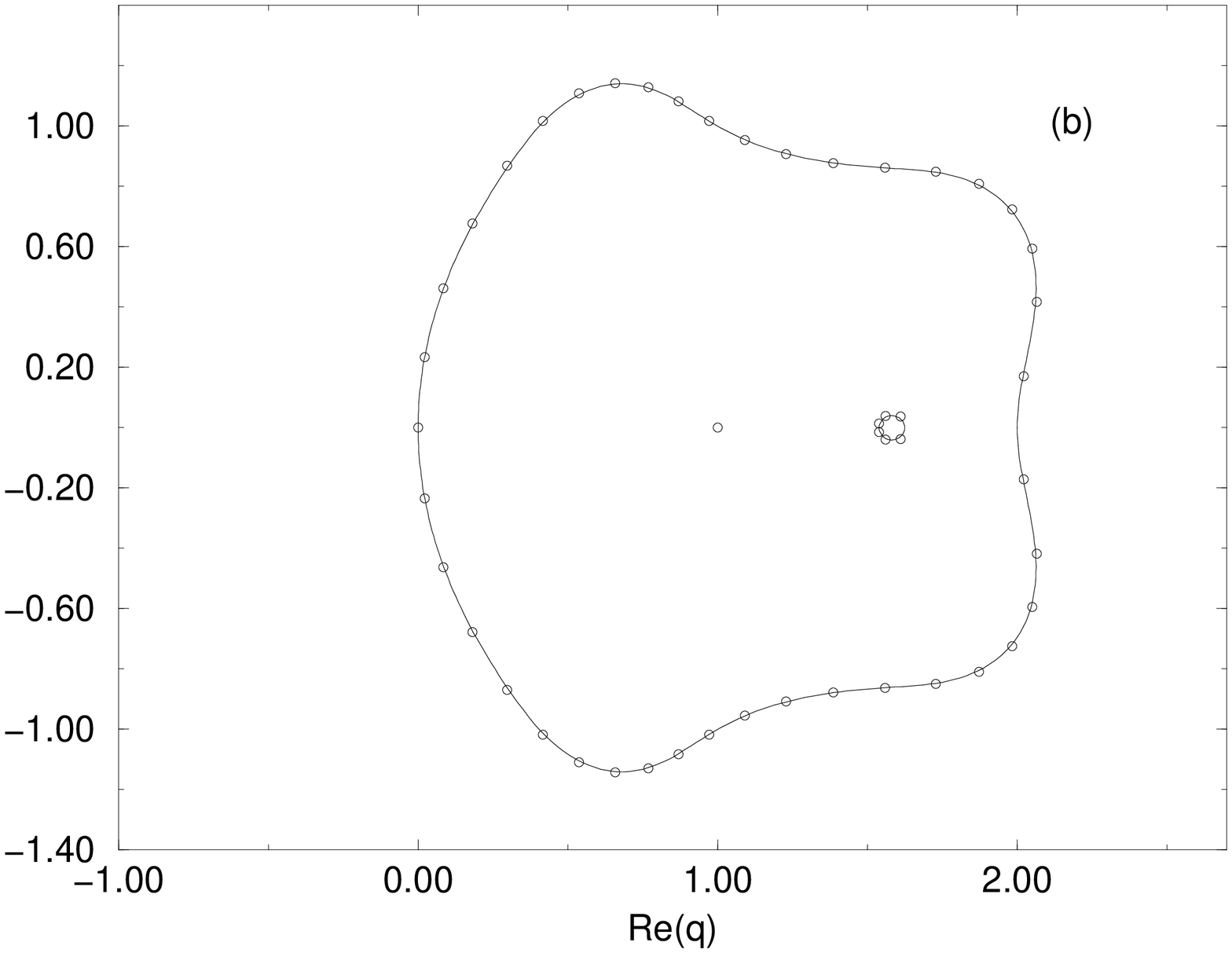}
\end{center}
\vspace{-2cm}
\caption{\footnotesize{Boundary ${\cal B}$ in the $q$ plane for $W$ function
for $\lim_{m \to \infty} G_{e_1,e_2,e_g,m}$ with $(e_1,e_2,e_g)=$
(a) (2,4,2), (b) (2,4,3). Chromatic zeros for (a) $m=8$ (i.e., $n=56$) and
(b) $m=6$ (i.e., $n=48$) are shown for comparison.}}
\label{nec24g23}
\end{figure}

\begin{figure}
\centering
\leavevmode
\epsfxsize=3.5in
\begin{center}
\leavevmode
\epsffile{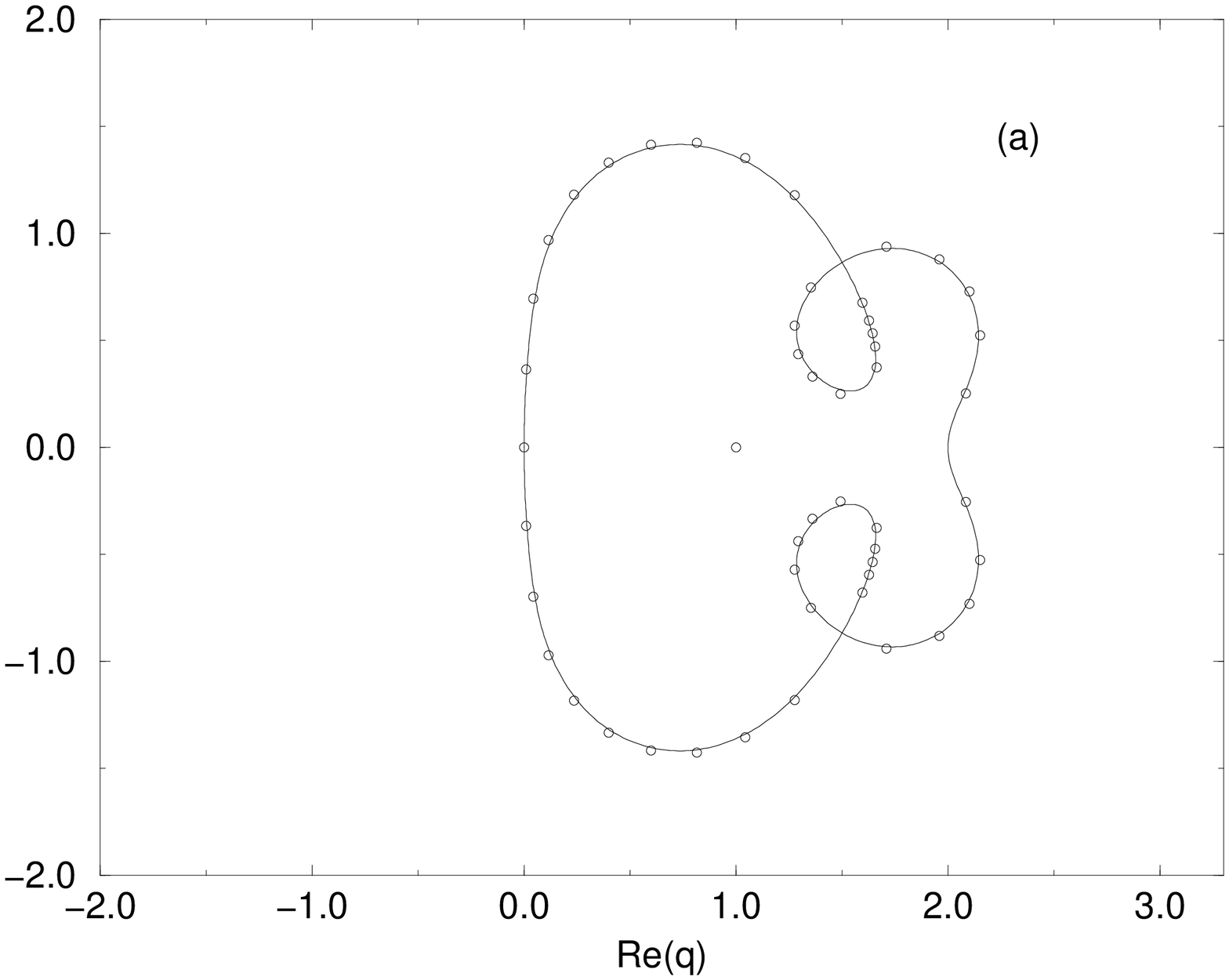}
\end{center}
\vspace{-4cm}
\begin{center}
\leavevmode
\epsfxsize=3.5in
\epsffile{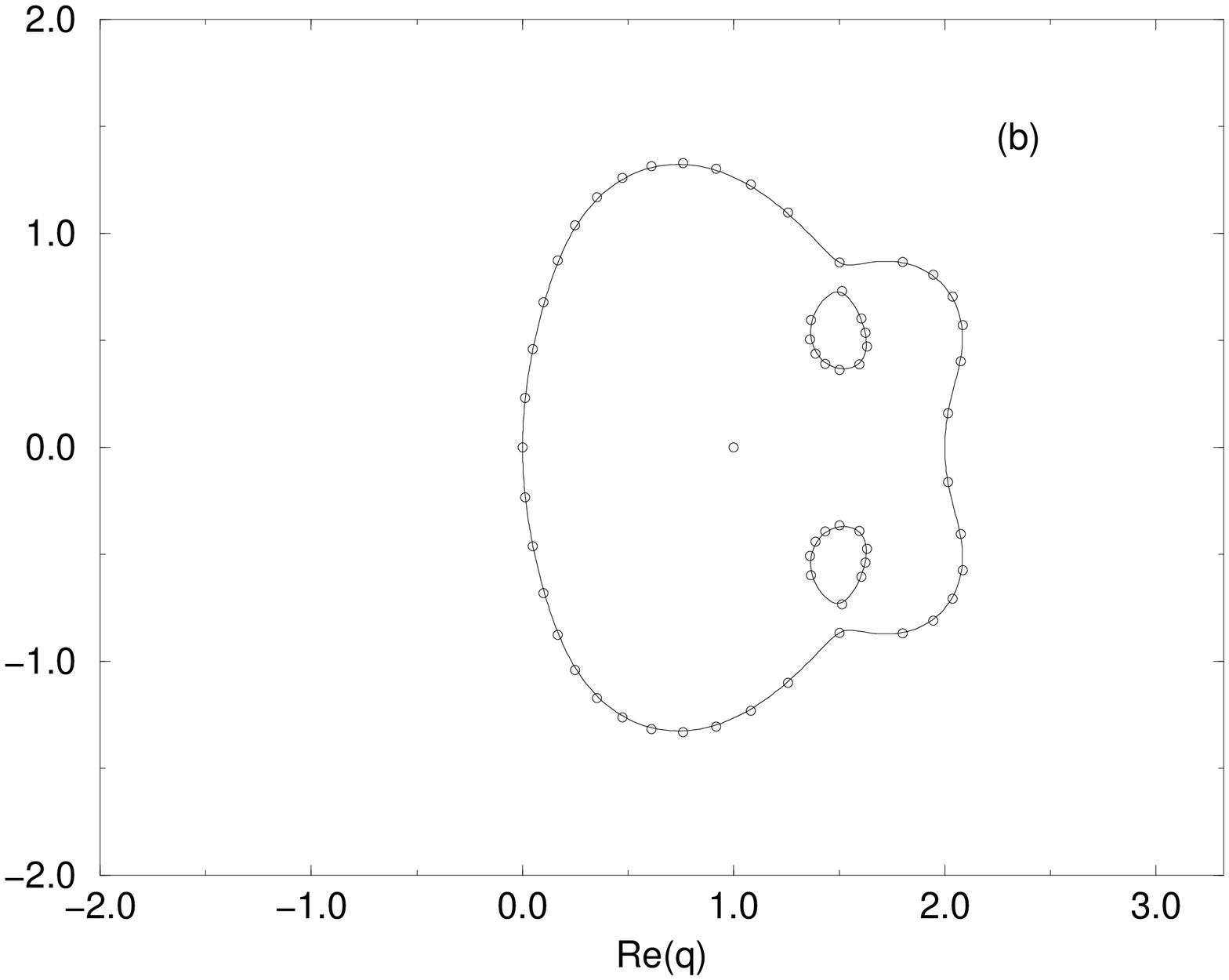}
\end{center}
\vspace{-2cm}
\caption{\footnotesize{Boundary ${\cal B}$ in the $q$ plane for $W$ function
for $\lim_{m \to \infty} G_{e_1,e_2,e_g,m}$ with $(e_1,e_2,e_g)=$
(a) (3,3,0), (b) (3,3,1). Chromatic zeros for $m=10$ (i.e., $n=50$ and
$n=60$ for (a) and (b)) are shown for comparison.}}
\label{nec33g01}
\end{figure}

\begin{figure}
\centering
\leavevmode
\epsfxsize=3.5in
\begin{center}
\leavevmode
\epsffile{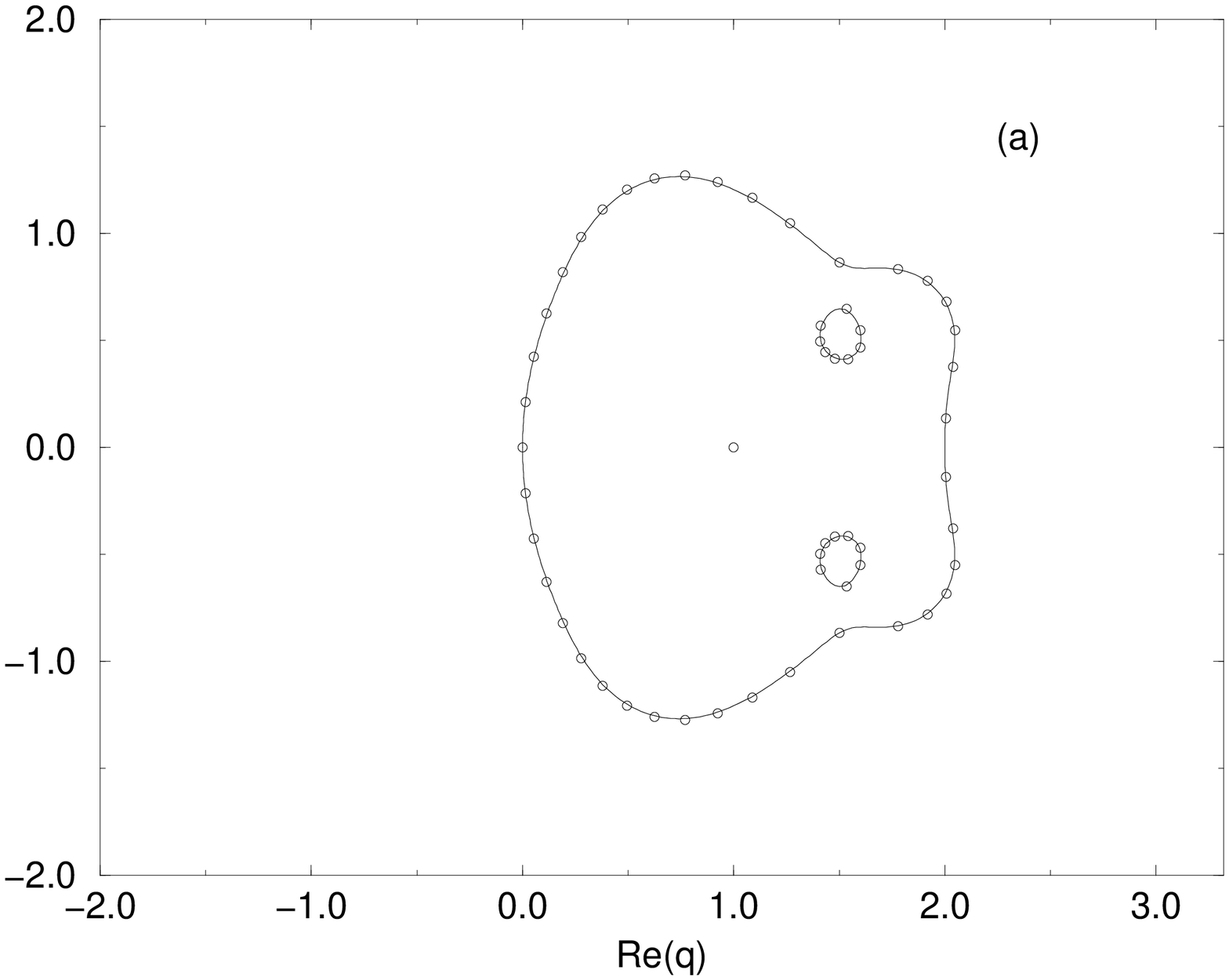}
\end{center}
\vspace{-4cm}
\begin{center}
\leavevmode
\epsfxsize=3.5in
\epsffile{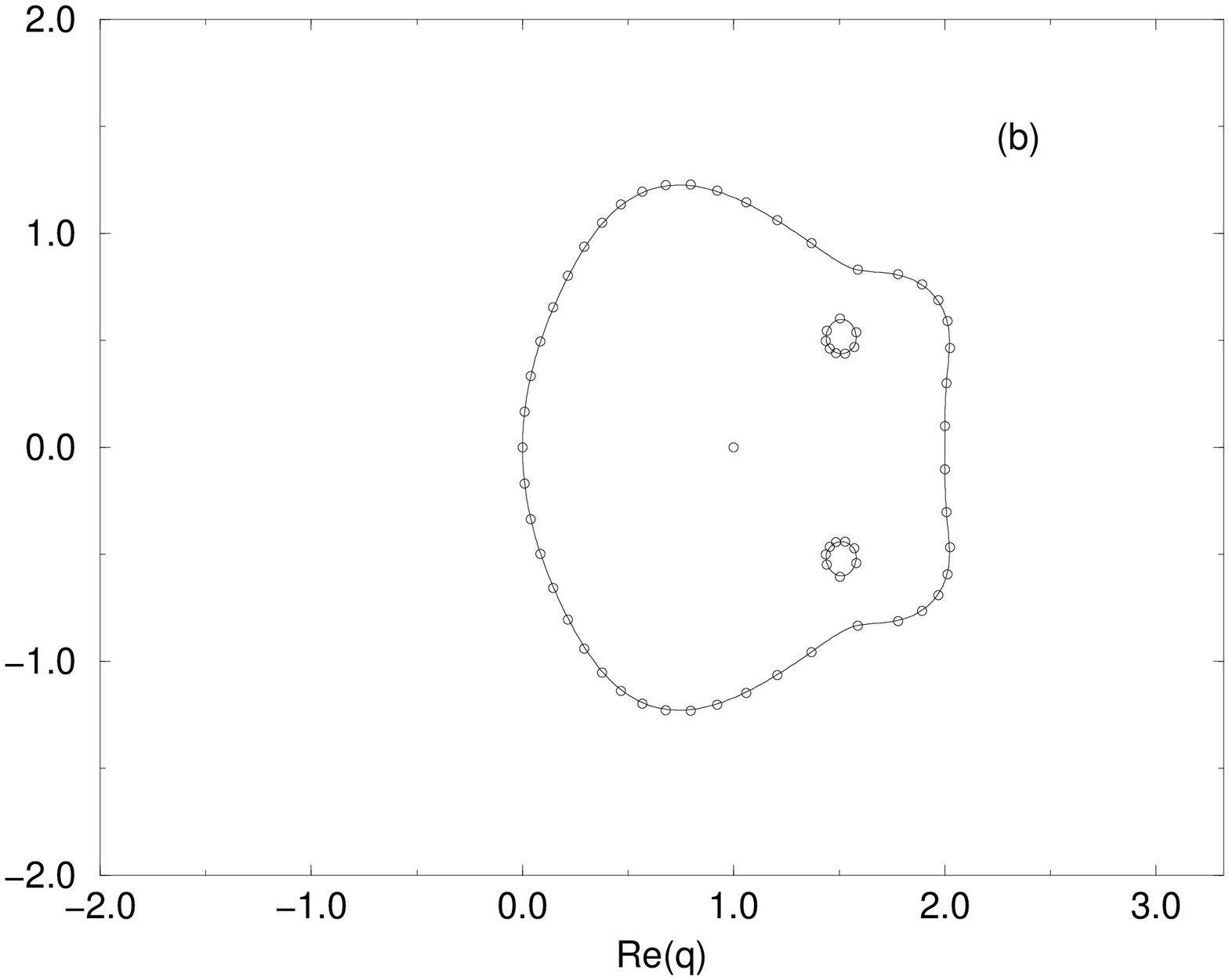}
\end{center}
\vspace{-2cm}
\caption{\footnotesize{Boundary ${\cal B}$ in the $q$ plane for $W$ function
for $\lim_{m \to \infty} G_{e_1,e_2,e_g,m}$ with $(e_1,e_2,e_g)=$
(a) (3,3,2), (b) (3,3,3). Chromatic zeros for $m=8$ (i.e., $n=56$) and
$n=64$ for (a) and (b)) are shown for comparison.}}
\label{nec33g23} 
\end{figure}

In Figs. (\ref{nec22g01})-(\ref{nec33g23}) we have also shown the chromatic
zeros of $G_{e_1,e_2,e_g,m}$ for large finite values of $m$ to compare with the
$m \to \infty$ accumulation set ${\cal B}$.  We observe first that the
chromatic zeros lie quite close to the asymptotic locus ${\cal B}$.  Second,
the density of chromatic zeros appears to vanish at multiple points, just as
was the case for multiple points on complex-temperature phase boundaries for
which we carried out exact calculations previously \cite{ih}. As one increases
$e_g$ above 0, so that new components of ${\cal B}$ pinch off, one still sees a
remnant of the reduction of density on the sides facing the location of the
multiple point that had existed for $e_g=0$.

\pagebreak

\pagebreak

\pagebreak

\pagebreak

\pagebreak

\section{Conclusions} 

In this paper we have given exact expressions for the chromatic polynomial
$P(G,q)$ and the resultant exponent of the ground state entropy for the Potts
antiferromagnet in the $n \to \infty$ limit, $W(\{G\},q)$ for families of
cyclic polygon chain graphs $G = G_{e_1,e_2,e_g,m}$.  We have studied several
types of limits yielding $n \to \infty$, viz., , $L_{e_g}$, $L_p$, and $L_m$,
the last of which yielded the most interesting results.  From these
calculations we have derived the continuous locus, ${\cal B}$, of
nonanalyticities of $W$, which is also the accumulation set of the zeros of the
chromatic polynomial in the $n \to \infty$ limit.  Various properties of this
locus were proved.  A comparison with the results for the open polygon chain
graph shows the important effect of global circuits.  The results of the
present study agree with and extend our earlier inferences concerning the locus
${\cal B}$, in particular, that a sufficient condition that in the $n\to
\infty$ limit the locus ${\cal B}$ separates the $q$ plane into two or more
regions is that the graph has a global circuit with $\lim_{n \to \infty}
\ell_{g.c.} = \infty$.

\vspace{10mm}

This research was supported in part by the NSF grant PHY-97-22101.

\vspace{10mm}

\section{Appendix}

In this appendix we present results for the families of cyclic 
polygon chain graphs with $(e_1,e_2)=(2,4)$ and (3,3), and, for each case, 
$0 \le e_g \le 3$.  

\newpage 

\pagebreak

\pagebreak

\pagebreak

\vfill
\eject


\begin{thebibliography}{99}

\bibitem{potts}{Potts, R. B. 1952 Proc. Camb. Phil. Soc. {\bf 48}, 106.}

\bibitem{wurev}{Wu, F. Y. 1982 Rev. Mod. Phys. {\bf 54}, 235;
Wu, F. Y. 1983 {\it ibid}. {\bf 55}, 315 (errata).}

\bibitem{al}{Aizenman, M. and Lieb, E. H. 1981 J. Stat. Phys. {\bf 24}, 279;
Chow, Y. and Wu, F. Y. 1987 Phys. Rev. {\bf B36}, 285.}

\bibitem{bl}{Birkhoff, G. D. and Lewis, D. C. 1946 Trans. Am. Math. Soc.
{\bf 60}, 355.}

\bibitem{rrev}{Read, R. C. 1968 J. Combin. Theory {\bf 4}, 52.}

\bibitem{rtrev}{Read, R. C. and Tutte, W. T. 1988 ``Chromatic Polynomials'',
in {\it Selected Topics in Graph Theory, 3}, (Academic Press, NY).}

\bibitem{w}{Shrock, R. and Tsai, S.-H. 1997 Phys. Rev. {\bf E55}, 5165.}

\bibitem{lieb}{Lieb, E. H. 1967 Phys. Rev. {\bf 162}, 162.}

\bibitem{bds}{Biggs, N. L., Damerell, R. M. and Sands, D. A. 1972
J. Combin. Theory B {\bf 12}, 123; Biggs, N. L., Meredith, G. H. 1976 
{\it ibid.}, B {\bf 20} 5.}

\bibitem{bkw}{Beraha, S., Kahane, J., and Weiss, N. 1980 J. Combin. Theory B
{\bf 28}, 52.}

\bibitem{bax}{Baxter, R. J. 1987 J. Phys. A {\bf 20}, 5241.}

\bibitem{read91}{Read, R. C. and Royle, G. F. 1991 in {\it Graph Theory,
Combinatorics, and Applications} (Wiley, NY), vol. 2, p. 1009.}

\bibitem{rw}{Read, R. C. and Whitehead, E. G., Discrete Math. in press.}

\bibitem{p3afhc}{Shrock, R. and Tsai, S.-H. 1997 J. Phys. A {\bf 30}, 495.}

\bibitem{wc}{Shrock, R. and Tsai, S.-H. 1997 Phys. Rev. {\bf E56}, 1342,
4111.}

\bibitem{wa}{Shrock, R. and Tsai, S.-H. 1997 Phys. Rev. {\bf E56}, 3935.}

\bibitem{ww}{Shrock, R. and Tsai, S.-H. 1997 Phys. Rev. {\bf E55}, 6791; 
Phys. Rev. {\bf E56}, 2733, 4111.}

\bibitem{wa3}{Shrock, R. and Tsai, S.-H. 1998 J. Phys. A {\bf 31}, 9641.}

\bibitem{strip}{Ro\v{c}ek, M., Shrock, R., and Tsai, S.-H. 1998 Physica
{\bf A252}, 505; Physica {\bf A259}, 367.}

\bibitem{hs}{Shrock, R. and Tsai, S.-H. 1998 Physica {\bf A259}, 315;
Phys. Rev. {\bf E58}, 4332, cond-mat/9808057.} 

\bibitem{wa2}{Shrock, R. and Tsai, S.-H. 1999 Physica {\bf A265}, 186.}

\bibitem{pg}{Shrock, R. and Tsai, S.-H. 1999 J. Phys. A Lett. {\bf 32} L195; 
ITP-SB-98-60.} 

\bibitem{alg}{Hartshorne, R. 1977 {\it Algebraic Geometry} (Springer, New
York).}

\bibitem{ih}{Matveev, V. and Shrock, R. 1995 J. Phys. A {\bf 28}, 4859; 
Matveev, V. and Shrock, R. 1996 Phys. Rev. {\bf E53}, 254.}

\end{thebibliography}
\end{document}